\documentclass[12pt,preprint]{aastex61}
\usepackage{epstopdf}
\usepackage{natbib}

\newcommand       \Teffqk      {\left(\frac{{T_{\rm eff}}}{1000{\rm K}}\right)}
\newcommand       \RV           {{R_V}}
\newcommand       \AV           {{A_V}}
\newcommand       \Teff         {T_{\rm {eff}}}
\accepted{by APJ 05-21-2018}

\begin{document}
\title{The Ultraviolet Extinction in the \emph{GALEX} Bands}


%

\author{Mingxu Sun}
\affiliation{Department of Astronomy,
               Beijing Normal University,
               Beijing 100875, China}
\author{B.~W. Jiang}
\affiliation{Department of Astronomy,
               Beijing Normal University,
               Beijing 100875, China}
\author{He Zhao}
\affiliation{Department of Astronomy,
               Beijing Normal University,
               Beijing 100875, China}

\author{Jian Gao}
\affiliation{Department of Astronomy,
               Beijing Normal University,
               Beijing 100875, China}

\author{Shuang Gao}
\affiliation{Department of Astronomy,
               Beijing Normal University,
               Beijing 100875, China}

\author{Mingjie Jian}
\affiliation{Department of Astronomy,
               Beijing Normal University,
               Beijing 100875, China}
\affiliation{Department of Astronomy,
             The University of Tokyo,
             7-3-1 Hongo, Bunkyo-ku, Tokyo 113-0033, Japan}
\author{Haibo Yuan}
\affiliation{Department of Astronomy,
               Beijing Normal University,
               Beijing 100875, China}

\correspondingauthor{B.~W. Jiang}
\email{bjiang@bnu.edu.cn}

%
%
%
%
%
%
%
%
%
%
%
%
%
%
%

\begin{abstract}
Interstellar extinction in ultraviolet is the most severe in comparison with optical and infrared wavebands and a precise determination plays an important role in correctly recovering the ultraviolet brightness and colors of objects. By finding the observed bluest colors at given effective temperature and metallicity range of dwarf stars,  stellar intrinsic colors, $C^0_{\rm B,V}$, $C^0_{\rm NUV,B}$, $C^0_{\rm FUV,B}$ and $C^0_{\rm FUV,NUV}$, are derived according to the stellar parameters from the LAMOST spectroscopic survey and photometric results from the \emph{GALEX} and APASS surveys. With the derived intrinsic colors, the ultraviolet color excesses are calculated for about 25,000 A- and F-type dwarf stars. Analysis of the color excess ratios yields the extinction law related to the \emph{GALEX} UV bands: $E_{{\rm NUV,B}}$/$E_{{\rm B,V}} = 3.77$,
      $E_{{\rm FUV,B}}$/$E_{{\rm B,V}} = 3.39$,
      $E_{{\rm FUV,NUV}}$/$E_{{\rm B,V}} = -0.38$. The results agree very well with previous works in the $NUV$ band and in general with the extinction curve derived by \citet{1999PASP..111...63F} for $\RV=3.35$.
\end{abstract}

\keywords{ISM: dust, extinction --- ultraviolet: ISM --- ultraviolet: stars}

\section{Introduction}
Interstellar extinction rises steeply towards ultraviolet (UV), in particular around 2175{\AA} where a strong bump occurs. \citet{1990ARA&A..28...37M} estimated the extinction at 2000{\AA} to be about three times that at the visual 5000{\AA} for an average $\RV=3.1$ (defined as the ratio between the absolute extinction $\AV$ in the $V$ band and the color excess $E(B-V)$ in $B-V$). The high extinction makes the UV bands very appropriate  to investigate the dust properties in very diffuse regions where the extinction in visual bands becomes insignificant. For example, the extinction at high Galactic latitude, the essential region to study external galaxies, is very small. \citet{2013ApJ...771...68P} pointed out that errors in dust extinction can have disastrous effects on determination of cosmological parameters.  Moreover, the UV extinction can best constrain the dust species. The bump around 2175{\AA} is generally ascribed to molecule-sized carbonaceous grains, but no conclusion has been drawn on whether PAH particles or graphite grains are the carriers \citep{2003ARA&A..41..241D}.

The study of UV extinction began only after the space observation became practical because our atmosphere blocks all the cosmic UV radiation. Both spectroscopic and photometric data are taken to study the UV extinction law. The spectroscopy provides the possibility to obtain continuous extinction curve.    \citet{1972ARA&A..10..197B} summarized the results on the UV extinction based on the Orbiting Astronomical Observatory-2 with a wavelength down to 2000{\AA}. The bump around 2175{\AA} stood out clearly in the average extinction curve from 14 stars and the variation is prominent in both the continuum and the bump extinction towards a few sightlines. With its launch in 1978, the IUE (International Ultraviolet Explorer) satellite \citep{1978Natur.275..372B, 1978Natur.275..377B} collected a wealth of spectral data in the wavelength range $\sim$ 1170-3200{\AA} with a resolution of $\sim$ 6{\AA}, which formed the basis of abundant studies of the UV extinction. Among those studies, \citet{1990ApJS...72..163F} analyzed a sample of 78 stars and developed an analytical fitting method to the UV extinction curve. The fitting generally includes three components as a function of the wavenumber: (1) a linear background, (2) a Lorentzian-like Drude profile and (3) a far-UV curvature term. Combining the IUE data with the FUSE \citep[Far Ultraviolet Spectroscopic Explorer,][]{2000ApJ...538L...1M} observation between 905 and 1195{\AA}, \citet{2009AIPC.1135..110G}  investigated the extinction towards 75 stars covering a full UV wavelength range from 905  {\AA} to 3300  {\AA}. They also found significant difference in the strength of the far-UV rise and the width of the 2175 {\AA} bump. In a word, the extinction in UV is found to rise steeply with a bump around 2175  {\AA} and vary with sightlines based on the spectroscopic observations of several tens of stars.

On the photometric side, the early UV-dedicated Astronomical Netherlands Satellite \citep{1975A&A....39..159V} performed a 5 channel photometry at central wavelengths of approximately 1550, 1800, 2200, 2500, and 3300  {\AA} covering the 2175 {\AA}. \citet{1985ApJS...59..397S} made use of the ANS data and derived the UV interstellar extinction excesses for 1415 stars with spectral types B7 and earlier, which turns out to be the largest sample for studying the UV extinction thanks to the advantage of large-scale by photometry over spectroscopy. Their results are important in the general direction of stars for which the extinction curve-shape is unknown. Such stars are numerous because spectroscopy was performed only in very limited number of sightlines. Nevertheless, it was only bright (mostly visual magnitude $<$10), low-latitude (Galactic latitude $< 30\degr $ ) and early-type stars that \citet{1985ApJS...59..397S} analyzed. The situation for other environments, e.g. middle and late spectral-type or high-latitude stars, is not clear.

\emph{GALEX}, the Galaxy Evolution Explorer, performed the ever largest  survey in two UV bands, the \emph{FUV} (1344-1786 {\AA}) and the \emph{NUV} (1771-2831 {\AA}) band. The All-Sky Imaging Survey (AIS) covered an area of 22,080 deg$^2$ with a depth of $\sim 20/21$ mag (\emph{FUV}/\emph{NUV} in the AB system) for more than 200 million measurements \citep{2014AdSpR..53..900B}. This provides a huge database to calculate the UV interstellar extinction for millions of stars and to study the UV interstellar extinction variation towards various sightlines. Based on the \emph{GALEX} database, \citet{2013MNRAS.430.2188Y} studied the average extinction in the \emph{GALEX}/\emph{FUV} and \emph{GALEX}/\emph{NUV} bands by using the standard pair technique in combination with the SDSS spectroscopic information. They took the average colors of the stars with low SFD98 extinction \citep{1998ApJ...500..525S} and similar stellar parameters as stellar intrinsic colors. In this work, we try to derive the UV color excesses for individual stars and to study the UV extinction law. Different from \citet{2013MNRAS.430.2188Y}, the intrinsic color index is derived for individual stars from the measured stellar parameters instead of pair method, and we use a spectroscopic database from the LAMOST survey, which is much larger than the SDSS stellar spectral database.

\section{Data Preparation}\label{DATA}

The route of our method consists of (1) determination of the \emph{GALEX}/UV band-related intrinsic colors of stars, $C_{\rm B,V}^0$ (i.e.$(B-V)_0$) , $C_{\rm NUV,B}^0$, $C_{\rm FUV,B}^0$ and $C_{\rm FUV,NUV}^0$, (2) calculation of the color excess, $E_{\rm B,V}$ (i.e.$C_{\rm B,V}-C_{\rm B,V}^0$), $E_{\rm NUV,B}$, $E_{\rm FUV,B}$ and $E_{\rm FUV,NUV}$,  from the derived intrinsic colors and observed colors $C_{\rm B,V}$ (i.e.$(B-V)_{\rm observed}$) , $C_{\rm NUV,B}$, $C_{\rm FUV,B}$, and (3) derivation of the ratios of color excesses $E_{\rm NUV,B}/E_{\rm B,V}$, $E_{\rm FUV,B}/E_{\rm B,V}$ and $E_{\rm FUV,NUV}/E_{\rm B,V}$.  This method was originally developed by \citet{2014ApJ...788L..12W} to study the near-infrared extinction law and then applied by \citet{2016ApJS..224...23X} to the mid-infrared bands. It was proved to be able to obtain a high-precision extinction. This method combines the photometric and spectroscopic information of stars. In this work, the photometric data are taken from the \emph{GALEX} survey for the UV bands and from the APASS survey for the visual bands, and the spectroscopic data are from the LAMOST survey.%

\subsection{Photometric data: \emph{GALEX} and APASS}

The essential catalog we used is the \emph{GALEX} GR/6+7 \citep{2014AdSpR..53..900B} \footnote{\url{http://dolomiti.pha.jhu.edu/uvsky}} in two UV bands, i.e. \emph{FUV} ($\lambda_{{\rm eff}}=$1528 {\AA}, 1344-1786 {\AA}) and \emph{NUV} ($\lambda_{{\rm eff}}= $2310 {\AA}, 1771-2831 {\AA}).  Although \emph{GALEX} carried out two photometry surveys -- AIS (the All-Sky Imaging survey) and MIS (the Medium-depth Imaging Survey), only the data from the AIS survey is available in the newest release and used in our work. The AIS survey observed 28,707 fields covering a unique area of 22,080 square degrees with a typical depth of  20/21mag (\emph{FUV}/\emph{NUV}, in the AB mag system). In total, there are 71 million sources and the detections in the \emph{NUV} band exceeds significantly in the \emph{FUV} band.
We chose the photometry accuracy to be better than 0.20 mag in the \emph{NUV} band and 0.30 mag in the \emph{FUV} band  to make a compromise between the size of the sample and the quality of photometry. 

The UV bands are supplemented with the visual  $B$ and $V$ bands in order to compare with the color excess in $B-V$, for which the APASS (AAVSO Photometric All-Sky Survey) \citep{2009AAS...21440702H, 2014CoSka..43..518H}\footnote{\url{https://www.aavso.org/apass}} is selected. The APASS survey was conducted in five filters, Johnson $B$ and $V$ plus Sloan $g'$, $r'$, $i'$.  The catalog now contains photometry for 60 million objects in about 99$\%$ of the sky \citep{2016yCat.2336....0H}.  The catalog we used is APASS DR9. The limiting magnitude is about 19 mag in the $r'$ band ($\thicksim$10$\sigma$), with an astrometric accuracy of $\thicksim$0.1 arcsec \citep{2014IAUS..298..310L}. \citet{2014MNRAS.443.1192C} and \citet{2014IAUS..298..310L} concluded that its flux calibration with respect to the SDSS photometry produces a photometric accuracy of better than 2\% for a single frame and $\thicksim$2\%$-$3\% for the whole observation area. As the optical photometry usually is of high quality and the APASS catalog is large, the accuracy is required to be better than 0.05 mag in both $B$ and $V$ bands, and such requirement  brings no significant influence on the volume of the sample. 

\subsection{Spectroscopic Data: LAMOST}

LAMOST, the Large Sky Area Multi-Object Fiber Spectroscopic Telescope, is the largest spectroscopic survey in existence, containing spectra of millions stars \citep{2012RAA....12.1197C, 2014IAUS..298..428L}\footnote{\url{http://dr4.lamost.org/catalogue}}. The LAMOST/DR4 dataset we used provided stellar parameters, i.e. effective temperature $\Teff$, surface gravity $\log g$ and metal abundance $Z$, and their errors, for over 4 million stars. Limited by the location and the structure of the telescope, the observable sky is from -10$\degr$ to +90$\degr$ in stellar declination. 

The accuracy of $\Teff$ from the LAMOST spectroscopic survey, $\sigma$$_{\Teff}$$/$$\Teff$ is required to be better than 5\%, i.e. 250\,K at $\Teff=5000\,{\rm K}$.   After cross-matching with the \emph{GALEX}/UV catalog, it is found that very few giant stars from the spectroscopic surveys were measurable in the UV bands, which can be understood by their energy distribution mainly in the red. Therefore, only the dwarf stars are selected by requiring the surface gravity $\log g > 3.5$. As the quality of $\Teff$ depends on the templates and the spectral coverage, we picked up the dwarf stars with $\Teff$ in [6500, 8500]\,K, that is A-type and F-type. The stars with $\Teff < 6500$\,K, i.e. G-, K- and M-type dwarf stars, usually emit excess UV radiation that may come from eruptive chromospheric activity and make the determination of intrinsic color index quite uncertain. On the upper limit of $\Teff$, the uncertainties of stellar parameters increase significantly when $\Teff > 8500$\,K.
This sample is thus complementary to that of \citet{1985ApJS...59..397S} which calculated the UV excess of stars earlier than B7.

\subsection{Combination of photometric and spectroscopic data}

\begin{figure}
\centering
\centerline{\includegraphics[scale=0.5]{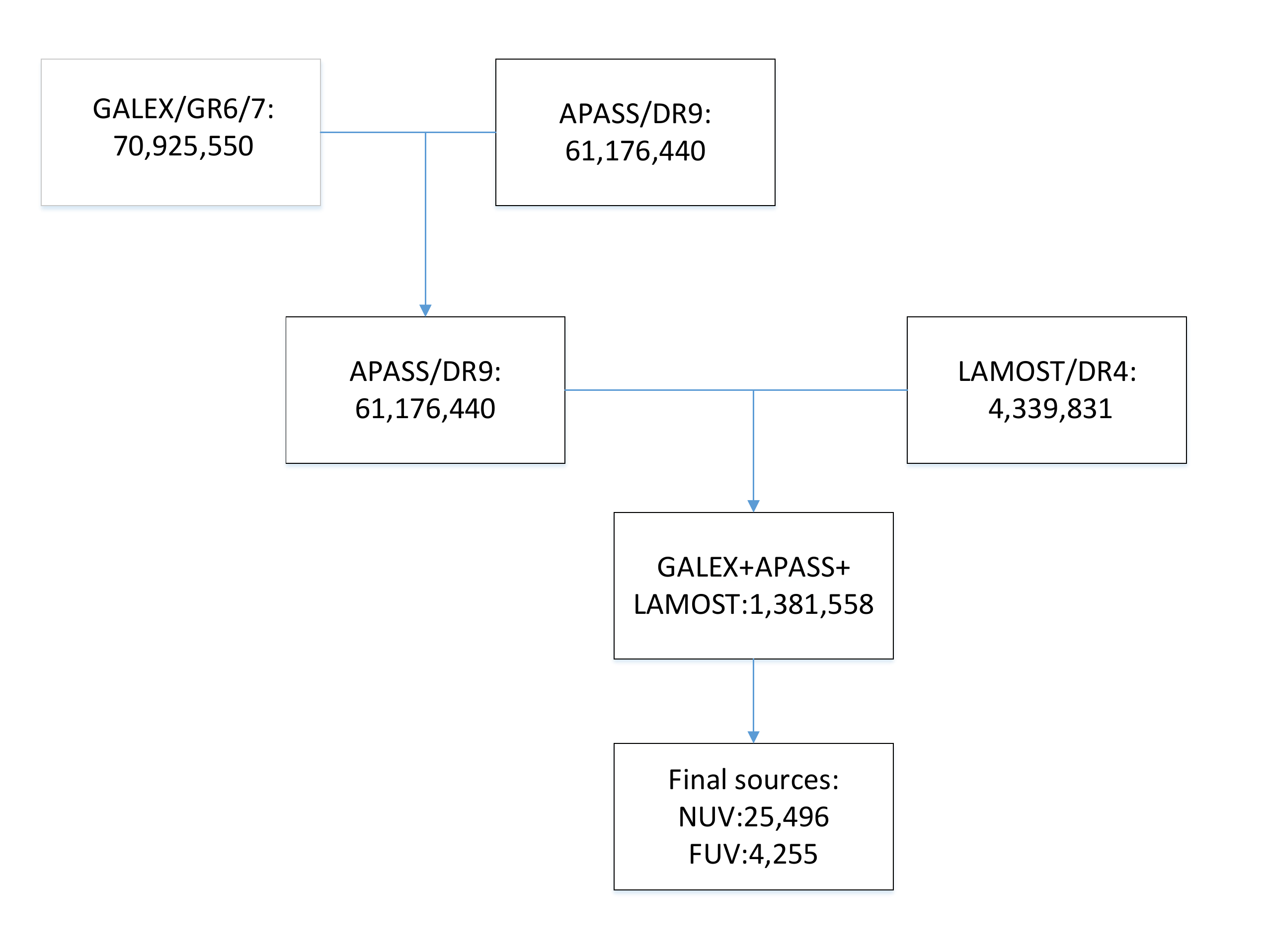}}
\caption{The number of the sources in and between the catalogs used.
\label{fig1}}
\end{figure}

The cross-match is carried out with a radius of 3$''$. The \emph{GALEX} DR6/7 and APASS DR9 catalogs are cross-matched first. This \emph{GALEX}-APASS catalog is further cross-identified with the LAMOST spectroscopic survey. The number of stars after cross identification and quality control is listed in Figure~\ref{fig1}. The final sample consists of 25,496 NUV and 4,255 FUV measurements. The huge number of stars lost from penultimate to ultimate boxes in Figure~\ref{fig1} is caused by the constraints on the parameters. The major factor is the effective temperature that there are only about 7.5\% stars with 6500\,K $< \Teff <$ 8500\,K in the LAMOST sample, which can be understood by the domination of low-mass and thus low-temperature (from about 3600\,K to 6500\,K) stars in the main sequence. The constraints on metallicity and surface gravity as well as the qualities of  stellar parameters and photometry further reduce the number of stars. 

\section{Determination of the intrinsic colors in the \emph{GALEX} UV bands}\label{method}

\subsection{Definition of the Blue Edge in the $\Teff$ versus $C^{0}_{\rm B,V}$ Diagram}

We followed the essential idea originally proposed by \citet{2001ApJ...558..309D} to determine stellar intrinsic colors. For a given spectral type, the observed bluest colors represent the intrinsic colors because such stars suffer either no or very little interstellar extinction. The prerequisite is that the sample actually includes such un-reddened stars. \citet{2014ApJ...788L..12W} and \citet{2016ApJS..224...23X} adopted this method to determine the relationship of intrinsic near- and mid-infrared colors with effective temperature for G-type giants based on the APOGEE spectroscopic survey \citep{2011AJ....142...72E, 2015ApJS..219...12A} \footnote{\url{http://www.sdss.org/surveys/apogee/}}. \citet{2017AJ....153....5J} used this method to systematically determine the infrared colors of normal stars of A-, F-, G-, K- and M-type with the stellar parameters from the LAMOST and RAVE surveys. \citet{2018ApJ...855...12Z} used this method to determine the distance to and the near-infrared extinction of the Monoceros supernova remnant. Here we apply this method to the visual and UV bands.

\subsubsection{Selection of the zero-reddening stars}
The zero-reddening stars are selected from the $\Teff$ versus $C_{\rm B,V}$ (the observed color index $B-V$) diagram, where a blue edge is clearly visible, and the stars redward of the blue edge experiences interstellar extinction, as shown in Figure~\ref{fig2}. Following previous studies \citep[e.g.][]{2017AJ....153....5J}, $\Teff$ is divided into some 200\,K-wide bins and the median color index of the 10\% bluest stars is taken as the intrinsic one of the bin  with more than 10 sources. The selection of 10\% actually differs from previous work that usually took 5\% such as in \citet{2017AJ....153....5J, 2016ApJS..224...23X, 2014ApJ...788L..12W}. This change is made to match the stellar model PARSEC in the intrinsic color $C_{\rm B,V}^0$, which will be illustrated in Section 3.4. In fact the final color excess ratio derived from linear fitting of two color excesses is little affected by the choice of percentage. The slope of linear fitting (see Section 4 for details) results in $E_{{\rm NUV,B}}$/$E_{{\rm B,V}}$ to be 3.75 for a choice of 5\%, 3.77 for 10\% and 3.77 for 15\%.  Afterwards, a polynomial fitting is performed to all the selected intrinsic colors which yields a relation of intrinsic color $C_{\rm B,V}^{0}$ with the effective temperature $\Teff$.  

The colors related to the UV bands are determined in a different way. The blue edge in the $\Teff$-$C_{\rm FUV,B}$/$C_{\rm FUV,NUV}$ diagrams looks very vague. One reason is that the \emph{FUV} sources are not numerous enough. The other reason is the large scatter in the color-$\Teff$ diagram, which may be intrinsic in the \emph{FUV} bands or caused by the relatively big uncertainty in the \emph{FUV} photometry (see Section 3.3 for the average errors of photometry in related bands). For the $NUV$ band, the method same as for determining $C_{\rm B,V}^{0}$ might be used since a blue edge is visible in the $\Teff$-$C_{\rm NUV,B}$ diagram. However, the blue-edge method is not adopted because we are not very certain to which percentage of the bluest stars should be taken when no good calibration in this band is available. Thus, the zero-reddening stars are selected from their color excess in $B-V$ to secure the reliability for the UV-related bands. In principle, the zero color excess in $B-V$, $E_{\rm B,V} \equiv C_{\rm B,V} - C_{\rm B,V}^{0}$, implies zero color excess in any other color indexes. The zero-reddening stars are chosen to be those whose $C_{\rm B,V}$ deviates less than 0.07 mag from the curve of $C_{\rm B,V}^{0}$ with $\Teff$.  The choice of 0.07 mag  considers the combined error of photometry in the $B$ and $V$ bands. Although the quality control set an upper limit of 0.05 mag for both bands, the average error is about 0.03 mag and 0.07 mag corresponds to approximately 3-sigma. These stars are taken as the zero-reddening objects in determining the colors $C_{\rm NUV,B}^{0}$, $C_{\rm FUV,B}^{0}$ and $C_{\rm FUV,NUV}^{0}$. $\Teff$ is also divided into the 200\,K-wide bins and the median color index of zero-reddening stars is taken as the intrinsic one of the bin with more than 10 sources. In addition, $\Teff$ is required to be bigger than 7000\,K in the case of $C_{\rm FUV,B}^{0}$ and $C_{\rm FUV,NUV}^{0}$  because the stars with $\Teff < 7000\,K$ then have very small extinction and are of no help in improving the precision of the FUV  extinction. As expected, the selected stars concentrate along the blue edge area (see Figures \ref{fig3} and \ref{fig4}).

\subsubsection{Division of the metallicity}

The influence of metallicity on stellar colors is expected because metallicity would affect the opacity. While the effect is small in the infrared \citep{2017AJ....153....5J}, it already becomes visible in the optical \citep{2005ApJ...626..446R}. Since metals have many lines in the UV range, the influence of metallicity should be even more serious in the UV bands.
Considering the uncertainty of measurement and the size of the sample, the metallicity  [M/H] is divided into three groups by a step of 0.25: [-0.625,-0.375], [-0.375,-0.125] and [-0.125,0.125]. There are stars with metallicity outside these ranges, but they are too few to derive a reliable relation and thus dropped off.

\begin{figure}
\centering
\centerline{\includegraphics[scale=1.2]{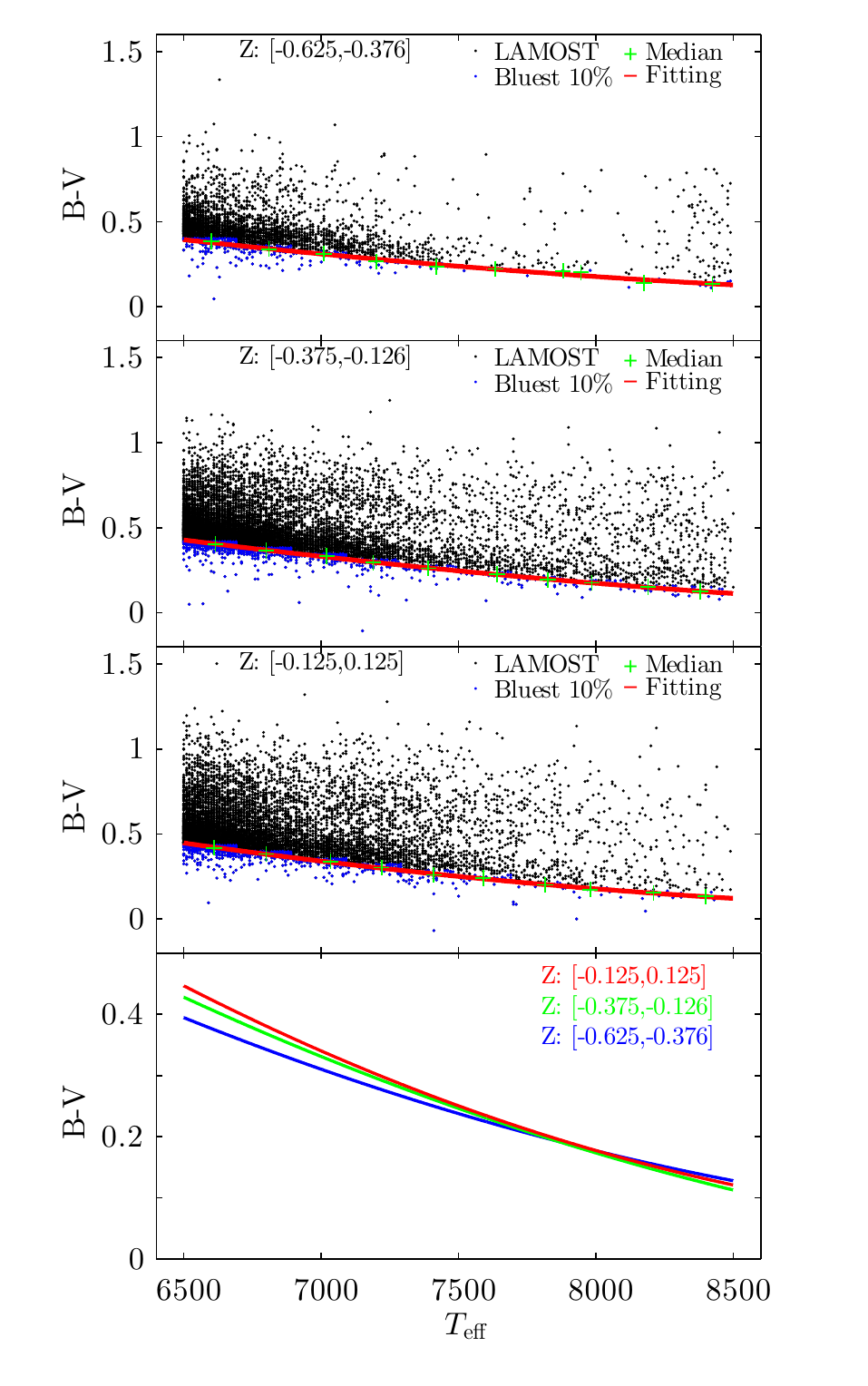}}
\caption{Determination of the relation of the color index $C^{0}_{\rm B,V}$ with $\Teff$ for the sample stars from the LAMOST survey. The  bluest 10\% stars are denoted by blue dots and their median colors by green crosses. The fitting curve of $C^{0}_{\rm B,V}$ with $\Teff$ is shown by a red solid line. The top three panels are for the metallicity range in [-0.625,-0.375], [-0.373,-0.125] and [-0.125,0.125] respectively. The final panel displays specifically the influence of metallcity on the relation of  $C^{0}_{\rm B,V}$ with $\Teff$.
\label{fig2}}
\end{figure}

\begin{figure}
\centering
\centerline{\includegraphics[scale=1.2]{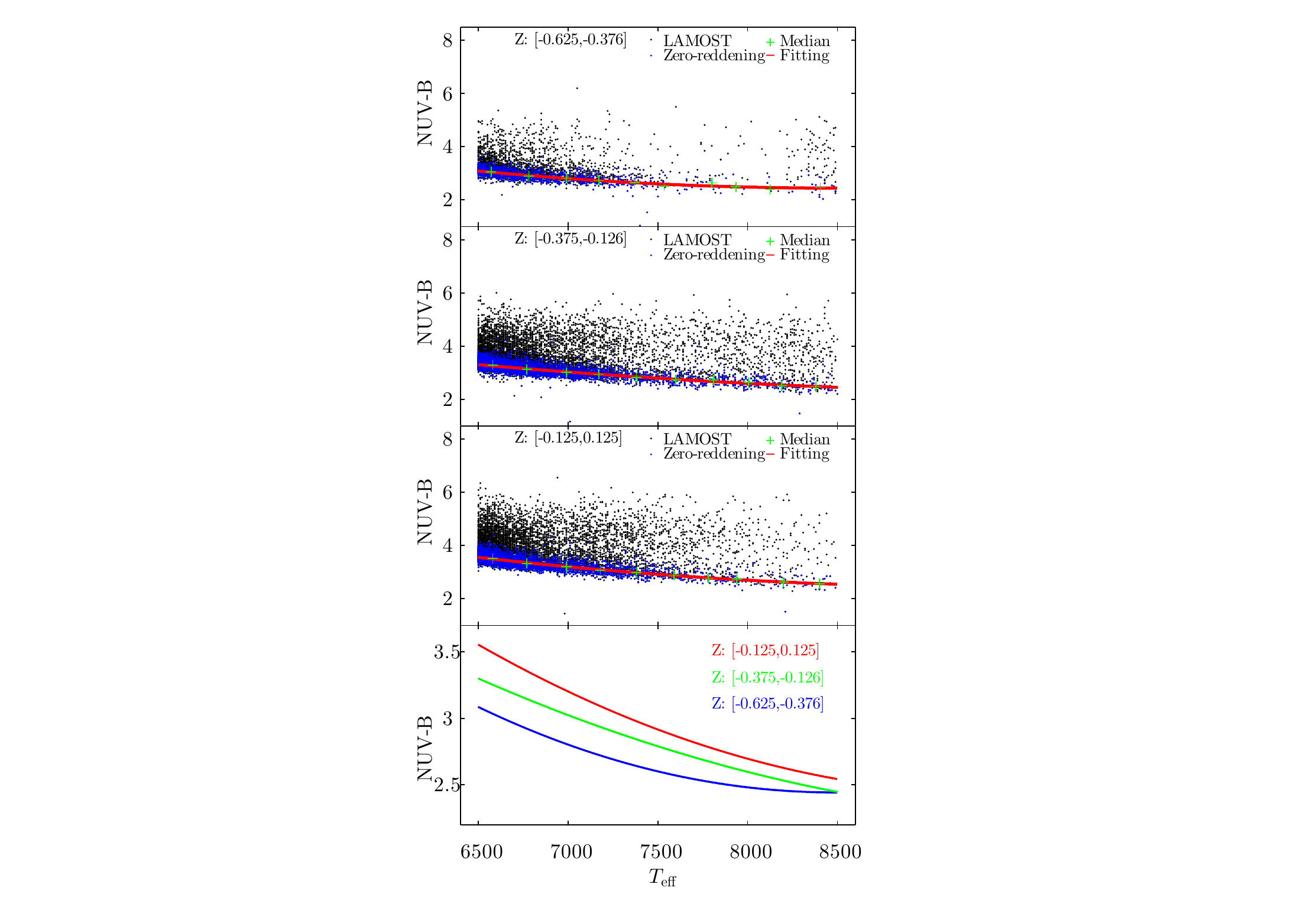}}
\caption{The same as Figure \ref{fig2}, but for the color index $C^{0}_{\rm NUV,B}$ with $\Teff$. Besides, the zero-reddening stars are selected from their location in Figure \ref{fig2} (see the text for details).
\label{fig3}}
\end{figure}

\begin{figure}
\centerline{\includegraphics[scale=1.2]{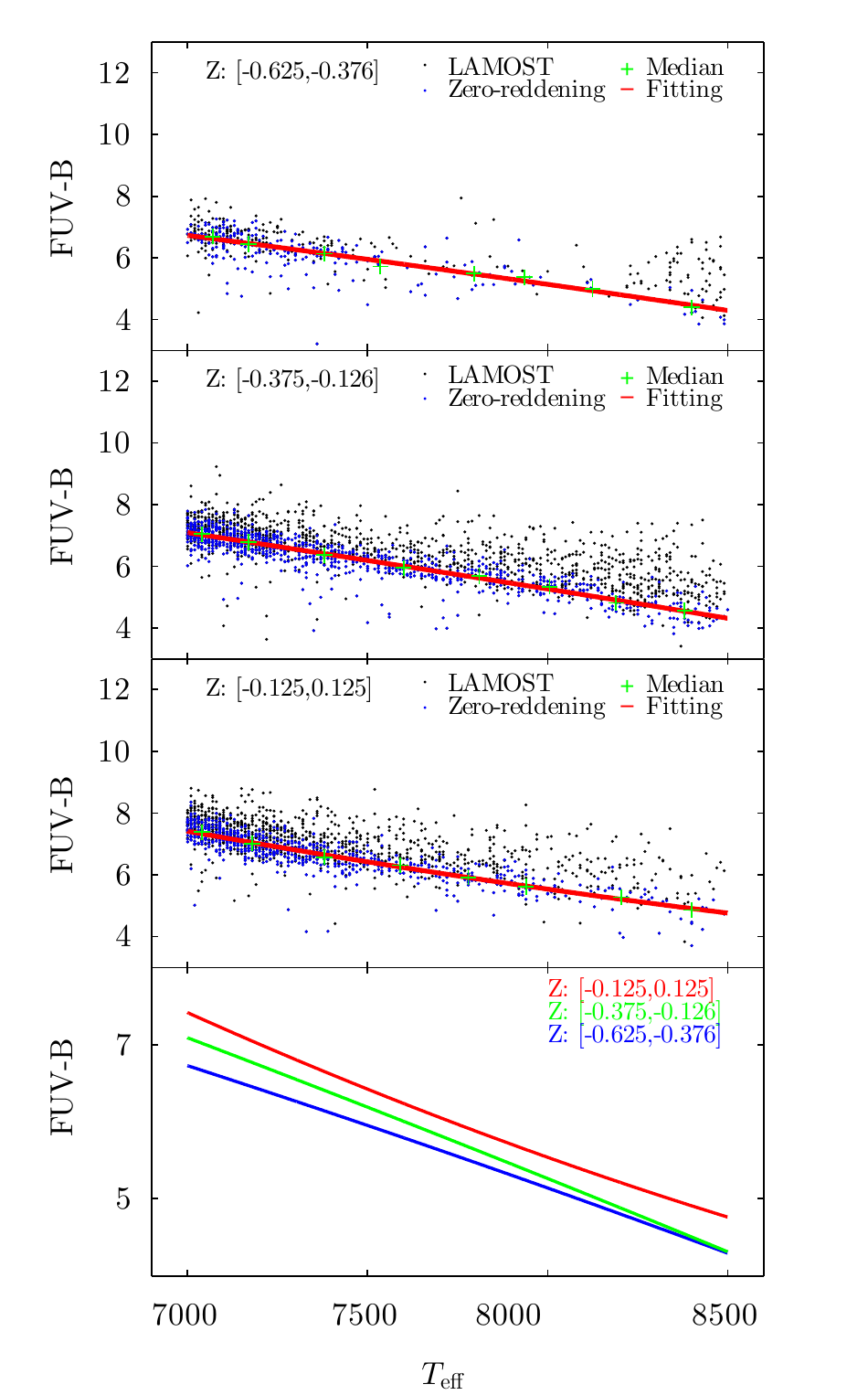}}
\caption{The same as Figure \ref{fig3}, but for the color index $C^{0}_{\rm FUV,B}$ with $\Teff$. In addition, the lower limit of $\Teff$ is increased to 7000\,K (see the text for details).
\label{fig4}}
\end{figure}

The relation of intrinsic colors with $\Teff$ in given metallicity ranges are shown in the last panel of Figure~\ref{fig2}, Figure~\ref{fig3}, Figure~\ref{fig4} and Figure~\ref{fig5}. As expected, the colors become redder with metallicity. The color indexes in the metallicity division of [-0.625, -0.375] at $\Teff>$8000\,K may be a little over-estimated due to the scarcity of stars in the given ranges (see the top panels of Figure \ref{fig2} and Figure \ref{fig3}). The average increase for $Z$ from -0.5 to 0 is 0.016 in $C^{0}_{\rm B,V}$, 0.305 in $C^{0}_{\rm NUV,B}$, 0.470 in $C^{0}_{\rm FUV,B}$, and 0.187 in $C^{0}_{\rm FUV,NUV}$. The metallicity effect is negligible in $C^{0}_{\rm B,V}$ but becomes significant in the \emph{GALEX}/UV bands.

\subsection{Relationships of $C^{0}_{\rm NUV,B}$, $C^{0}_{\rm FUV,B}$, $C^{0}_{\rm FUV,NUV}$  with  $\Teff$ }

The relation of $C^{0}_{\rm \lambda_1,\lambda_2}$ with $\Teff$ for the given metallicity division is derived by a quadratic function fitting of the zero-reddening stars, i.e.:

\begin{equation}\label{BV_Teff}
C^0_{\rm{\lambda_1,\lambda_2}}=a_0+a_1\times{\Teffqk}+a_2\times{\Teffqk}^2
\end{equation}

\begin{figure}
\centering
\centerline{\includegraphics[scale=1.2]{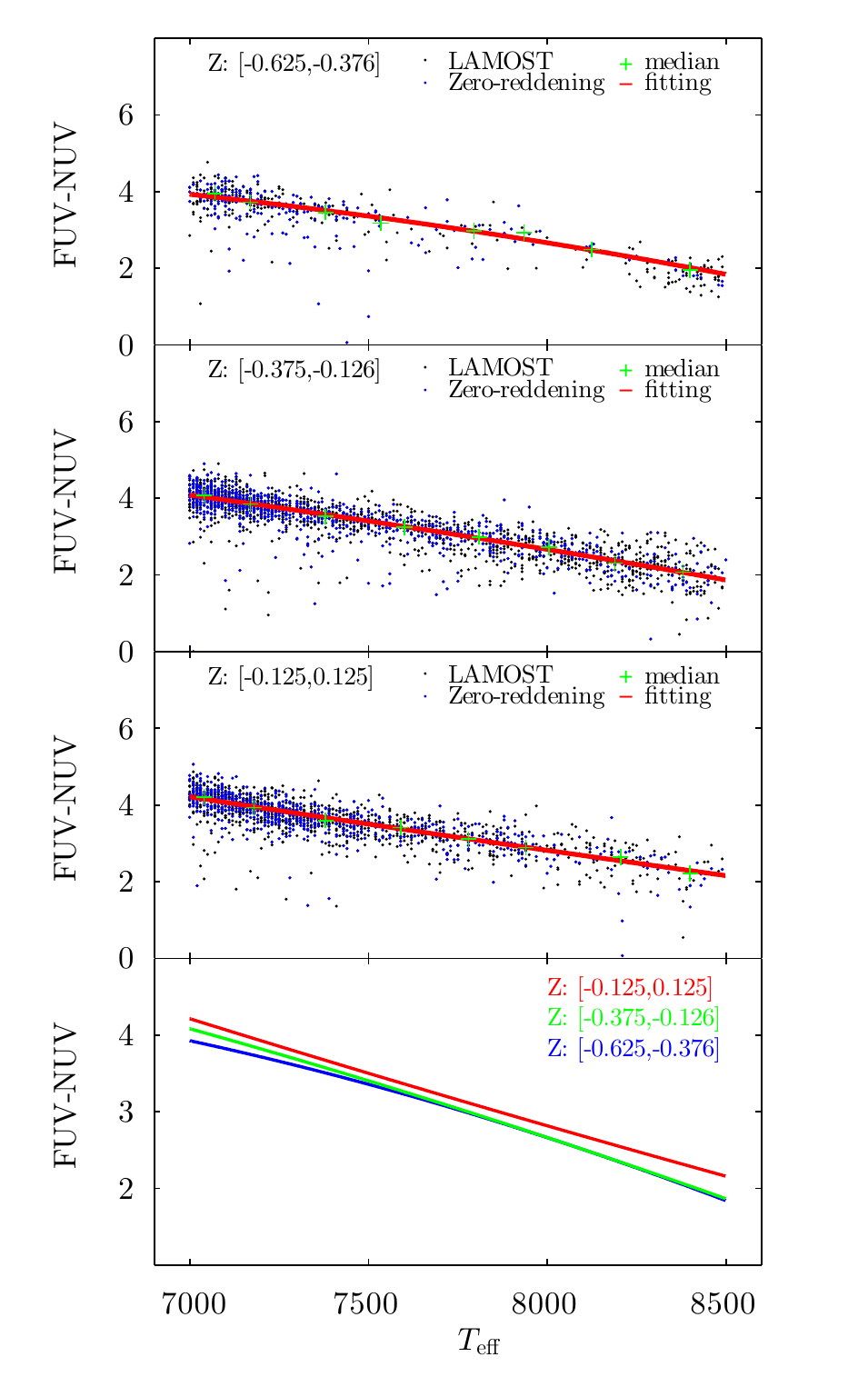}}
\caption{The same as Figure \ref{fig4}, but for the color index $C^{0}_{\rm FUV,NUV}$ with $\Teff$.
\label{fig5}}

\end{figure}

The coefficients are listed in Table\,\ref{table1}. The fitting curves for different metallicity range are decoded by red line in  Figure~\ref{fig2}, Figure~\ref{fig3}, Figure~\ref{fig4} and  Figure~\ref{fig5} for $C^{0}_{\rm B,V}$, $C^{0}_{\rm NUV,B}$, $C^{0}_{\rm FUV,B}$ and $C^{0}_{\rm FUV,NUV}$.

\begin{figure}
\centering
\centerline{\includegraphics[scale=1]{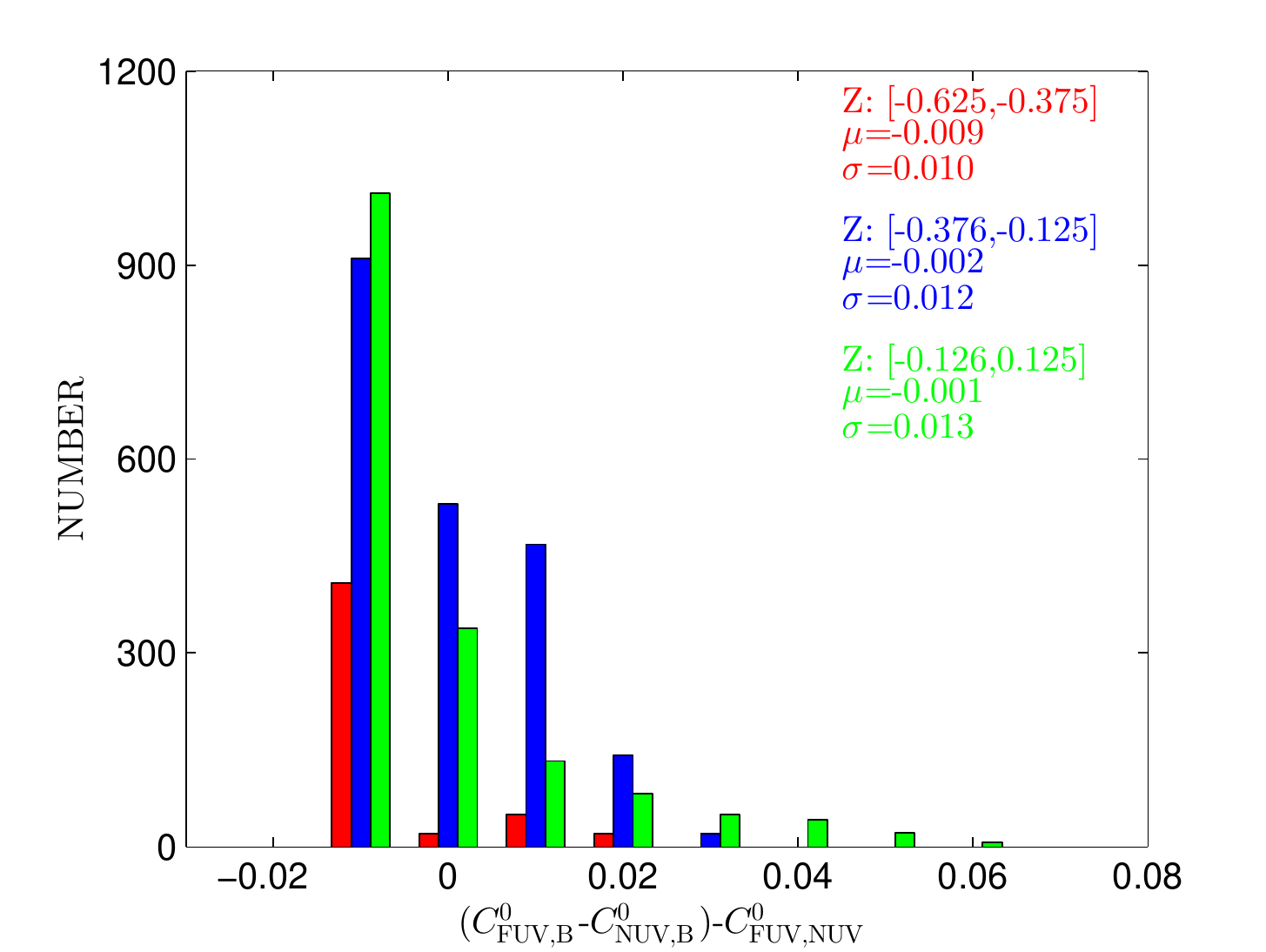}}
\caption{Difference of the intrinsic color index $C^{0}_{\rm FUV,NUV}$  derived directly by the blue-edge method as shown in Figure \ref{fig5} and that derived by $C^{0}_{\rm FUV,B}$-$C^{0}_{\rm NUV,B}$.
\label{fig6}}
\end{figure}

The internal consistency is examined by comparing $C^{0}_{\rm FUV,NUV}$ and $C^{0}_{\rm FUV,B}-C^{0}_{\rm NUV,B}$ which should equal to each other if no error present. Figure~\ref{fig6} demonstrates that $C^{0}_{\rm FUV,NUV}$ determined directly by the blue edge method by Figure \ref{fig5} is very consistent with that indirectly inferred from $C^{0}_{\rm FUV,B}-C^{0}_{\rm NUV,B}$, and the difference is on the order of 0.01 mag and can be taken as the uncertainty of fitting since they use the same sample of zero-reddening stars. The $C_{\rm FUV,NUV}$ vs $\Teff$ diagram (Figure~\ref{fig5}) exhibits large scatter, which will be understood by the following analysis that the extinctions in these two bands are comparable so that the difference between the observed and intrinsic colors is too small. On the other hand, the indiscrimination of the zero-reddening stars from the others (c.f. Figure \ref{fig5}) indicates that the color excess $E_{\rm FUV,NUV}$ is insignificant.  


\subsection{Uncertainty of the intrinsic color indexes}

The derived intrinsic color indexes suffer the errors brought by photometry, stellar parameters and the method. For the stars used to trace the extinction, the mean and dispersion of the photometric errors are,  $0.029\pm0.012$ in $B$, $0.028\pm0.012$ in $V$, $0.034\pm0.030$ in $NUV$, $0.145\pm0.073$ in $FUV$, which transfers into the mean error of observed color index to be 0.042 in $C_{{\rm B,V}}$, 0.048 in $C_{{\rm NUV,B}}$, 0.150 in $C_{{\rm FUV,B}}$, and 0.150 in $C_{{\rm FUV,NUV}}$.

The uncertainty of intrinsic color index $C^{0}_{\lambda1\lambda2}$ brought by the error in the key parameter $\Teff$ is measured by the difference with $\Teff$ shifted by an amount of its error, i.e. if the error of $\Teff$ is $\sigma_{T_{\rm {eff}}}$, $C^{0}_{\lambda1\lambda2}$ is re-calculated at $\Teff\pm \sigma_{T_{\rm {eff}}}$, then $\bigtriangleup C^{0}_{\lambda1\lambda2}$ between $C^{0}_{\lambda1\lambda2}({T_{\rm {eff}}}) $ and $C^{0}_{\lambda1\lambda2}({T_{\rm {eff}}}\pm \sigma_{T_{\rm {eff}}})$ is taken as the uncertainty associated with $\Teff$.

Metallicity is another parameter to affect $C^{0}_{\lambda1\lambda2}$ in addition to $\Teff$. Though we have divided stellar metallicity into groups, there is still a range of metallicity in each group. The uncertainty of $C^{0}_{\lambda1\lambda2}$ associated with metallicity is measured by the difference derived from the analytical formulas for two adjacent metallicity groups.

The major uncertainty of the blue-edge method lies in the choice of the bluest percentage of stars with zero-reddening, since the fitting error is very small implied by Figure \ref{fig6}. The presently chosen percentage, 10\%, leads to the consistency with the stellar model PARSEC in $C^{0}_{\rm B,V}$. Replacing 10\% by 5\% as in previous works would systematically shift $C^{0}_{\rm B,V}$ to bluer, while replacing by 15\% would shift to redder. We assign the difference with the choice of 5\% and 15\% to the uncertainty associated with the method.

The uncertainties from stellar parameters and the method are comparable, which are all small in comparison with the photometric error. The mean values of uncertainties are taken  and shown in Table \ref{error_c0}. The summarized uncertainty amounts to 0.048 in $C^{0}_{\rm B,V}$, 0.136 in $C^{0}_{\rm NUV,B}$, 0.242 in $C^{0}_{\rm FUV,B}$, and 0.187 in $C^{0}_{\rm FUV,NUV}$.

\subsection{Comparison with the PARSEC model}

\begin{figure}
\centering
\centerline{\includegraphics[scale=0.8]{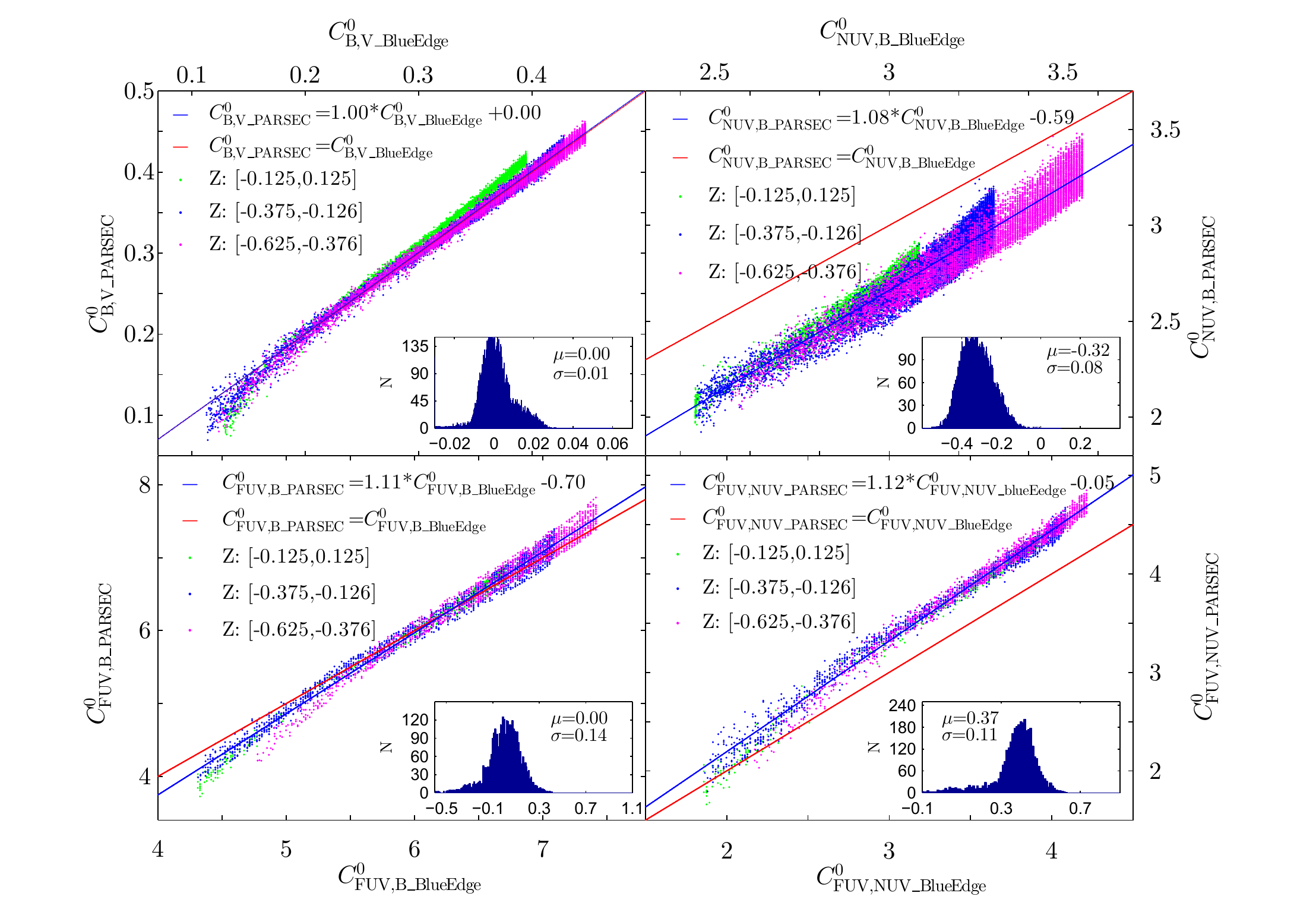}}
\caption{Comparison of the intrinsic colors derived by our blue-edge method (the $x$ axis) and the stellar model PARSEC (the $y$ axis). The linear fitting is denoted by blue line and the red line denotes the equal relationship, with the quantitative relations. The inset shows the distribution of the difference with its mean and standard deviation.
\label{fig7}}
\end{figure}

The intrinsic colors can also be derived from stellar models and they are compared. The PAdova TRieste Stellar Evolution Code (PARSEC) \citep{2012MNRAS.427..127B} is chosen to calculate the absolute magnitudes at given stellar parameters in the related bands (\emph{B}, \emph{V}, \emph{NUV} and \emph{FUV}) and thus the intrinsic colors between bands. The absolute magnitudes are obtained from interpolation between the nearest model points, the models fail for some stars with stellar parameters beyond the model grids, which may be caused by the error either in stellar parameters or in the models. Though this may be amended by choosing the closest model grid point, such extrapolation method can induce very uncertain values and is not applied \citep{2018ApJ...855...12Z}. Consequently, the PARSEC model obtains the intrinsic color $C^{0}_{\rm NUV,B}$ for 22,706 out of 25,496 stars, $C^{0}_{\rm FUV,B}$ for 3,613 out of 4,255 stars.
The intrinsic color of each star is calculated with the PARSEC model and compared with the result derived from the blue-edge method in Figure~\ref{fig7}. For $C^{0}_{\rm B,V}$, the model agrees very well with the blue-edge method with an average difference of 0.0015 mag, significantly smaller than the uncertainty of $C^{0}_{\rm B,V}$. If the 5\% bluest stars were taken, this difference would rise to 0.017 mag, which motivated us to take the 10\% bluest stars, since the PARSEC model in optical filters was carefully calibrated \citep{2012MNRAS.427..127B}. Still the model predicts a bluer $C^{0}_{\rm B,V}$  than the blue-edge method by about 0.05 mag at $C^{0}_{\rm B,V} < 0.15$, but there seems to be no right choice of percentage to bring about the agreement in the entire range of $\Teff$.  For $C^{0}_{\rm NUV,B}$, the model yielded a bluer index by $\sim -0.32$ on average.  On the other hand, the modelled $C^{0}_{\rm FUV,B}$ generally agrees with the blue-edge method and is only redder by about 0.001. $C^{0}_{\rm FUV,NUV}$ has the largest difference, i.e. about 0.37 mag on average redder than the blue-edge method. The inconsistent tendency of difference in $C^{0}_{\rm NUV,B}$ and $C^{0}_{\rm FUV,B}$ made us in dilemma that no agreeable way exists in shifting the blue-edge. Besides, there has no determination of stellar intrinsic colors in this UV wavelength range previously, it is difficult to judge whether the problem lies on the model due to the uncertain UV opacity. 

\subsection{Comparison with the very-low SFD extinction sightlines}

The  most widely used SFD extinction map \citep{1998ApJ...500..525S} provides another independent check of our determination of intrinsic colors. Based on this map, the zero-reddening stars are selected by the SFD sightlines that have color excess $E_{\rm B,V\_SFD} < 0.05$, which results in 7744 stars. Afterwards, the procedures are identical to the blue-edge method: stars grouped according to their metallicity,  zero-reddening stars fitted by a quadratic function, intrinsic colors calculated by the analytical formula from their effective temperature and metallicity group.

\begin{figure}
\centering
\centerline{\includegraphics[scale=1.2]{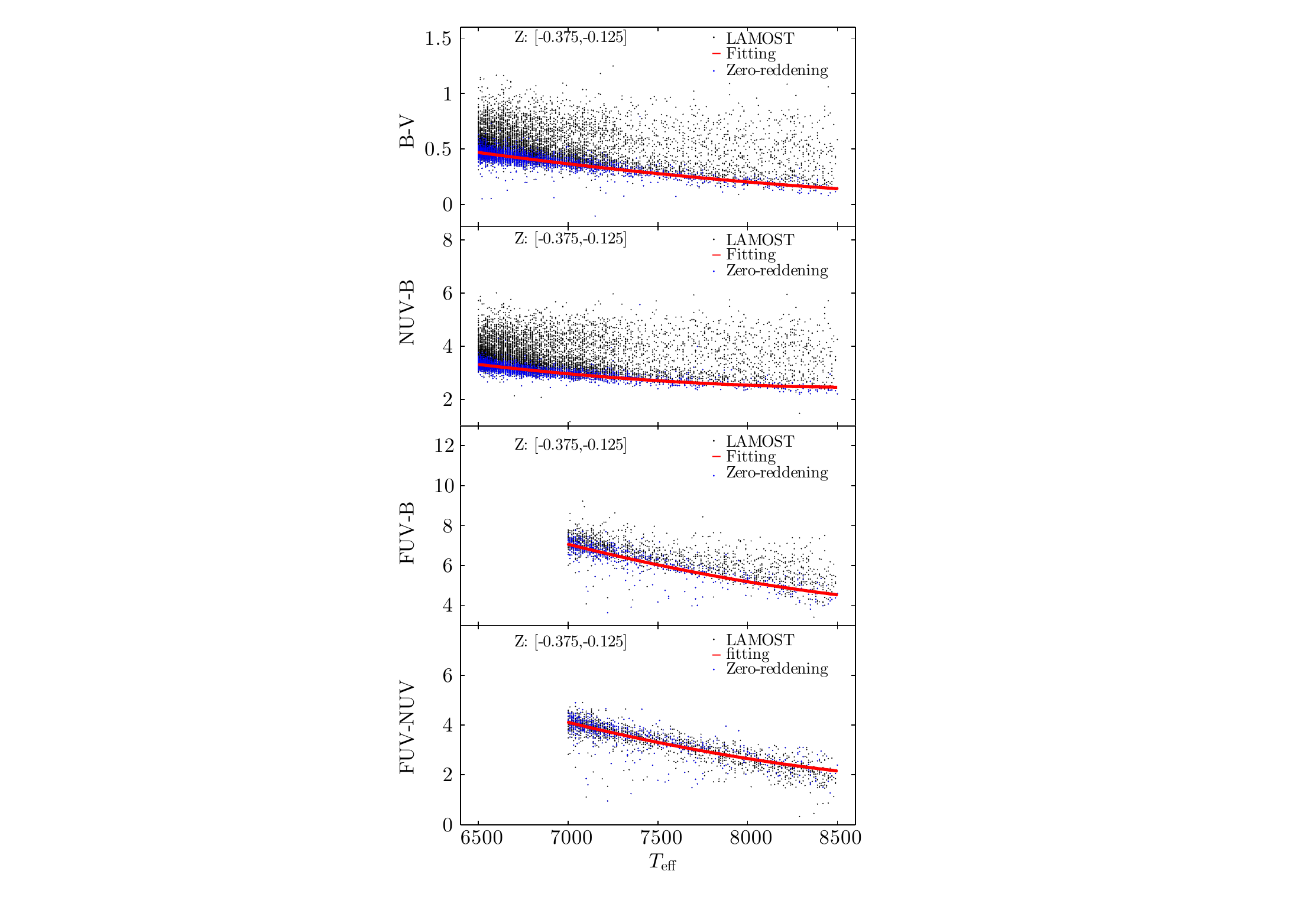}}
\caption{Determination of the relation of the color indexes $C^{0}_{\rm B,V}$, $C^{0}_{\rm NUV,B}$, $C^{0}_{\rm FUV,B}$  and $C^{0}_{\rm FUV,NUV}$ with $\Teff$ from the stars whose sightlines have  E(B-V) $<=$ 0.05 (denoted by blue dots) based on the SFD extinction map. The fitting curves are shown by a red solid line. Only the case for the metallicity range of [-0.375,-0.125] is shown.
\label{fig8}}
\end{figure}

\begin{figure}
\centering
\centerline{\includegraphics[scale=1]{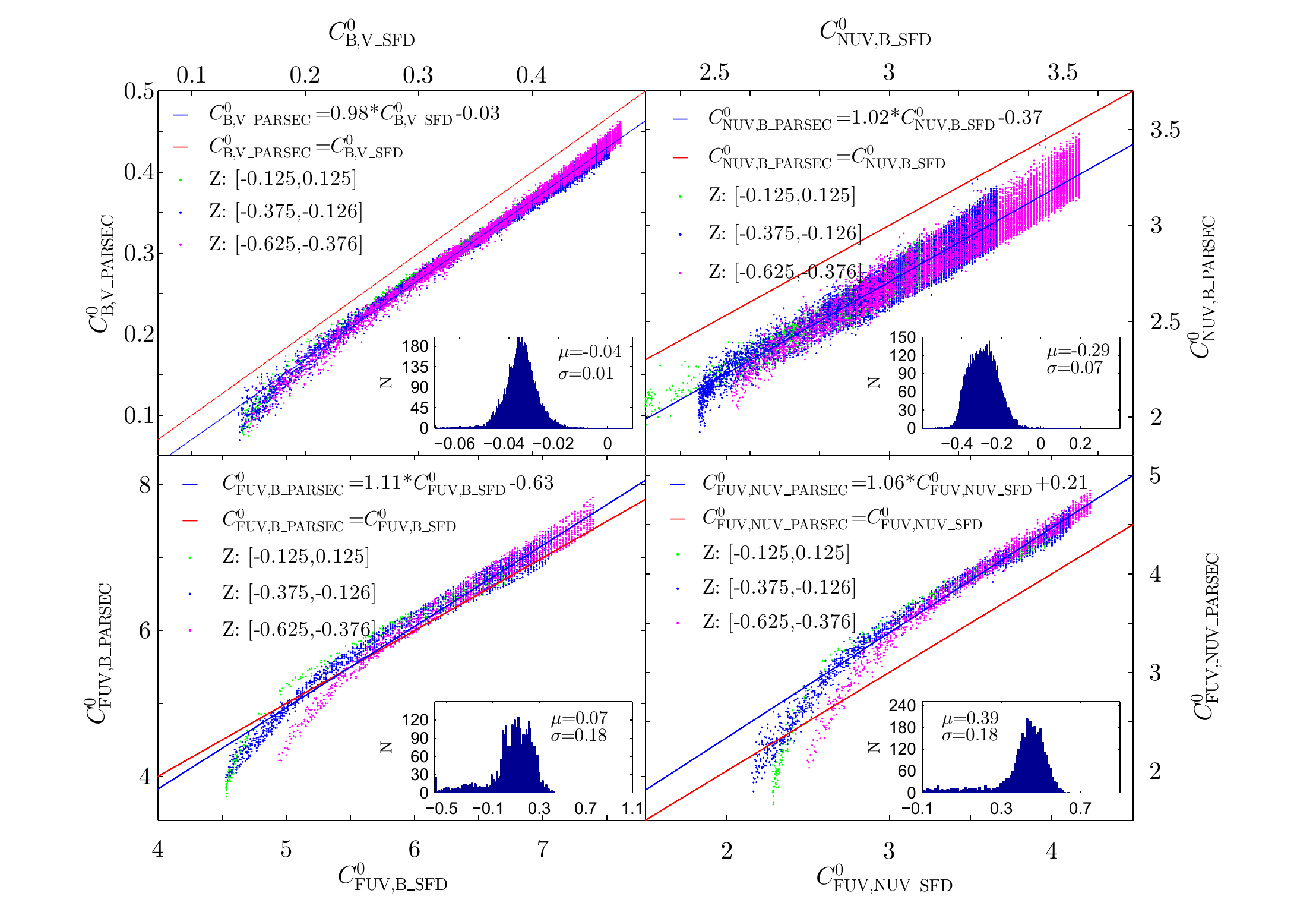}}
\caption{Comparison of the intrinsic color indexes derived from the SFD-selected zero-reddening stars (the $x$ axis) and the stellar model PARSEC (the $y$ axis). The linear fitting is denoted by the blue line and the red line denotes the equal relationship, with the quantitative relations. The inset shows the distribution of the difference with its mean and standard deviation.
\label{fig9}}
\end{figure}

Figure \ref{fig8} exemplifies the case for the metallicity division in [-0.375, -0.125] being the largest sub-sample for the LAMOST dataset.  First of all, the zero-reddening stars concentrate on the blue edge of the color-$\Teff$ diagram as expected. However, these stars whose sightlines with $E_{\rm B,V\_SFD} < 0.05$ disperse more widely than our zero-reddening stars selected by the 10\% blue-edge method. Consequently, the derived intrinsic color index $C^{0}_{\rm B,V}$ is redder by $\sim0.03$ mag, as shown in Figure \ref{fig9} for comparison with the PARSEC model. This can be understood by that these stars are not free of interstellar extinction. Though small, the interstellar extinction does affect. The average  $E_{\rm B,V}$ of the SFD zero-reddening sightlines is $ 0.033 \pm 0.010 $, which can completely account for the difference with the PARSEC model as well as with our blue-edge method. Similarly, $C^{0}_{\rm FUV,NUV}$ is bluer and can be understood by the extinction in $NUV$ exceeding in $FUV$ which will be shown later. Meanwhile, $C^{0}_{\rm NUV,B}$ is redder than the model by $\sim0.23$ mag, without no apparent reddening in comparison with blue-edge method.  Nevertheless, that $C^{0}_{\rm FUV,B}$ is bluer than the PARSEC model, which also occurs in the blue-edge method, owes an explanation.

\section{Results and Discussion}

\subsection{The color excess}

\begin{figure}
\centering
\centerline{\includegraphics[scale=1]{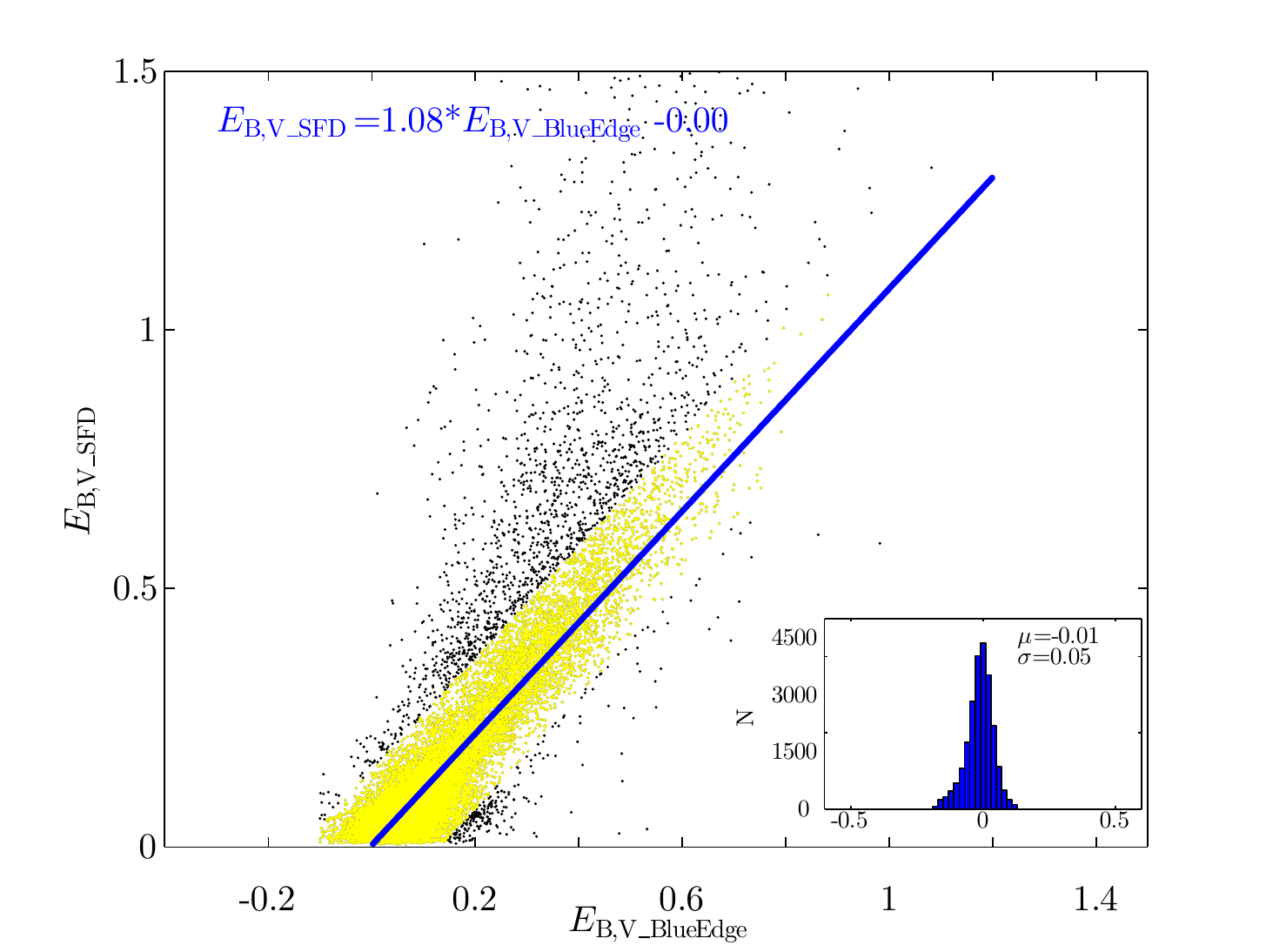}}
\caption{Comparison of the color excess $E(B-V)$ derived by our blue-edge work (the $x$ axis) and the SFD98 result (the $y$ axis). The inset shows the distribution of the difference with its mean and standard deviation.
\label{fig10}}
\end{figure}

The color excesses are straightforwardly calculated with the intrinsic colors derived and the observed colors at hands. The derived $E_{{\rm B,V}}$ is compared with that from SFD98 \citep{1998ApJ...500..525S} taken from the \emph{GALEX} catalog \citep{2014AdSpR..53..900B} in Figure \ref{fig10}. They agree with each other generally, meanwhile SFD98 yields systematically larger value at relatively large excess, which confirms the conclusion by various previous works that SFD98 over-estimated the extinction for being an integration over the whole sightline \citep[see e.g.][]{2005PASJ...57S...1D, 1999ApJ...512L.135A,2014MNRAS.443.1192C}. Because our sample are dwarf stars, most of them are nearby and may locate in front of the dust clouds responsible for the SFD extinction. Excluding the sightlines where $E_{{\rm B,V\_SFD}} > E_{{\rm B,V\_BlueEdge}}$ and some outliers, the match is almost perfect with no systematic difference and a dispersion of only 0.05 mag comparable to the uncertainty.

The color excesses, $E_{{\rm B,V}}$, $E_{{\rm NUV, B}}$, $E_{{\rm FUV, B}}$ and $E_{{\rm FUV,NUV}}$,  are obtained for all the sample stars. In total, there are 25,496 A- and F-type stars with NUV-related and 4,255 stars with FUV-related color excess from the LAMOST survey, which is an enormous increase  and supplement to the  1415 early-type stars by \citet{1985ApJS...59..397S}. The results are available through the CDS website.

\subsection{The color excess ratio}

The color excess ratios, $E_{{\rm NUV,B}}$/$E_{{\rm B,V}}$, $E_{{\rm FUV,B}}$/$E_{{\rm B,V}}$ and $E_{{\rm FUV,NUV}}$/$E_{{\rm B,V}}$ are derived from the linear fitting between color excesses. Although there should be no negative color excess theoretically, the measured color excess may be negative due to the uncertainty. Accounting for the upper limit of photometry error in the data selection, the cutoff of color excess is not zero but shifted to some negative value, about three times of the mean uncertainty. Specifically, $E_{\rm B,V} > -0.1$, $E_{{\rm NUV,B}} > -0.5$ and $E_{{\rm FUV,B}} > -1.0$ are required for further linear fitting. We use the  robust fitting and choose the bisquare weights. This method minimizes a weighted sum of squares where the weight given to each data point depends on how far the point is from the fitted line. The bisector linear fitting is another option, in which the bisecting line  of the Y vs X and X vs Y fits is determined. In principle, the bisector linear fitting works better when the uncertainties in X and Y are comparable and there is no true independent variable. With the errors calculated in Section 3.3 and  shown in Table~\ref{error_c0}, the color excesses bear the uncertainty of 0.064 in $E_{{\rm B,V}}$, 0.134 in $E_{{\rm NUV,B}}$, 0.279 in $E_{{\rm FUV,B}}$, and 0.233 in $E_{{\rm FUV,NUV}}$. The uncertainty in  $E_{{\rm B,V}}$ is apparently smaller than in other color excesses. Nevertheless, both the robust fitting with bisquare weights and the bisector fitting are tried to the data. Usually the goodness of fitting is evaluated by the residuals. The two methods have little difference in the goodness in terms of residuals. In our case, the other important factor to measure the goodness is the intercept. Since both color excesses should pass the zero point simultaneously, the intercept is expected to be zero in ideal situation. But due to the possible systematic shift in intrinsic color indexes, the linear fitting results in non-zero intercept, which would change the real color excess ratio. If the line is forced to pass the origin, the slope shall be changed. Therefore, the smaller the intercept, the better the fitting.
It is found that the two methods resulted in comparable residuals. However, the intercept\footnote{The intercepts are -0.13 and -0.16 for robust and bisector fitting respectively of $E_{{\rm NUV,B}}$ vs. $E_{{\rm B,V}}$, -0.09 and -0.19 for robust and bisector fitting respectively of $E_{{\rm FUV,B}}$ vs. $E_{{\rm B,V}}$.} is larger in the case of bisector fitting. Moreover, the internal consistency among three color excess ratios is better for the robust fitting method. Therefore, we decided to adopt the result from robust fitting with the bisquare weights.

\begin{figure}
\centering
\centerline{\includegraphics[scale=1]{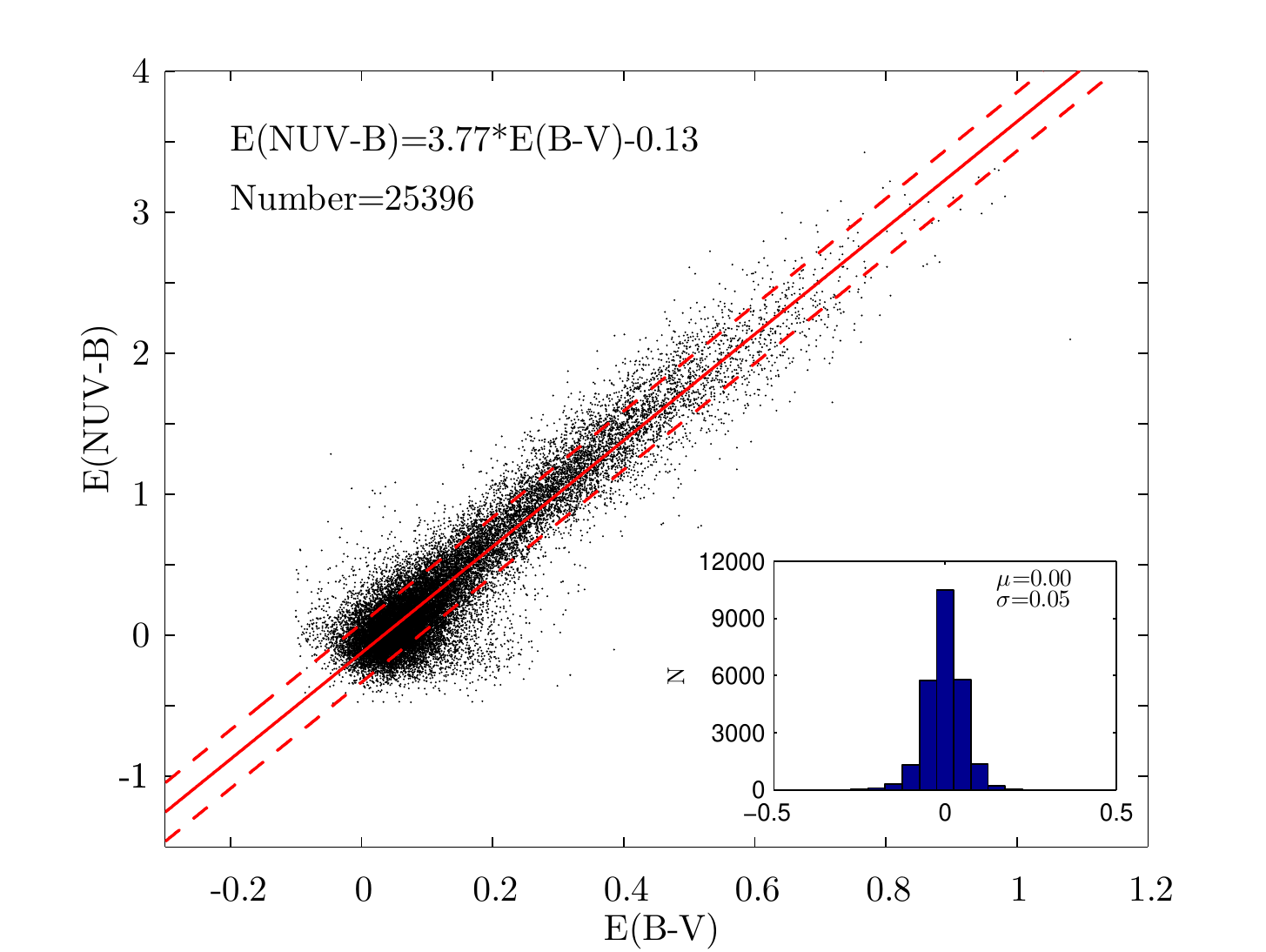}}
\caption{Linear fitting of the color excesses, $E_{{\rm NUV,B}}$ and $E_{{\rm B,V}}$ for the LAMOST stars. The inset shows the distribution of the residuals with its mean and standard deviation, where the residual is the perpendicular distance to the fitting line.
\label{fig11}}
\end{figure}

\begin{figure}
\centering
\centerline{\includegraphics[scale=1]{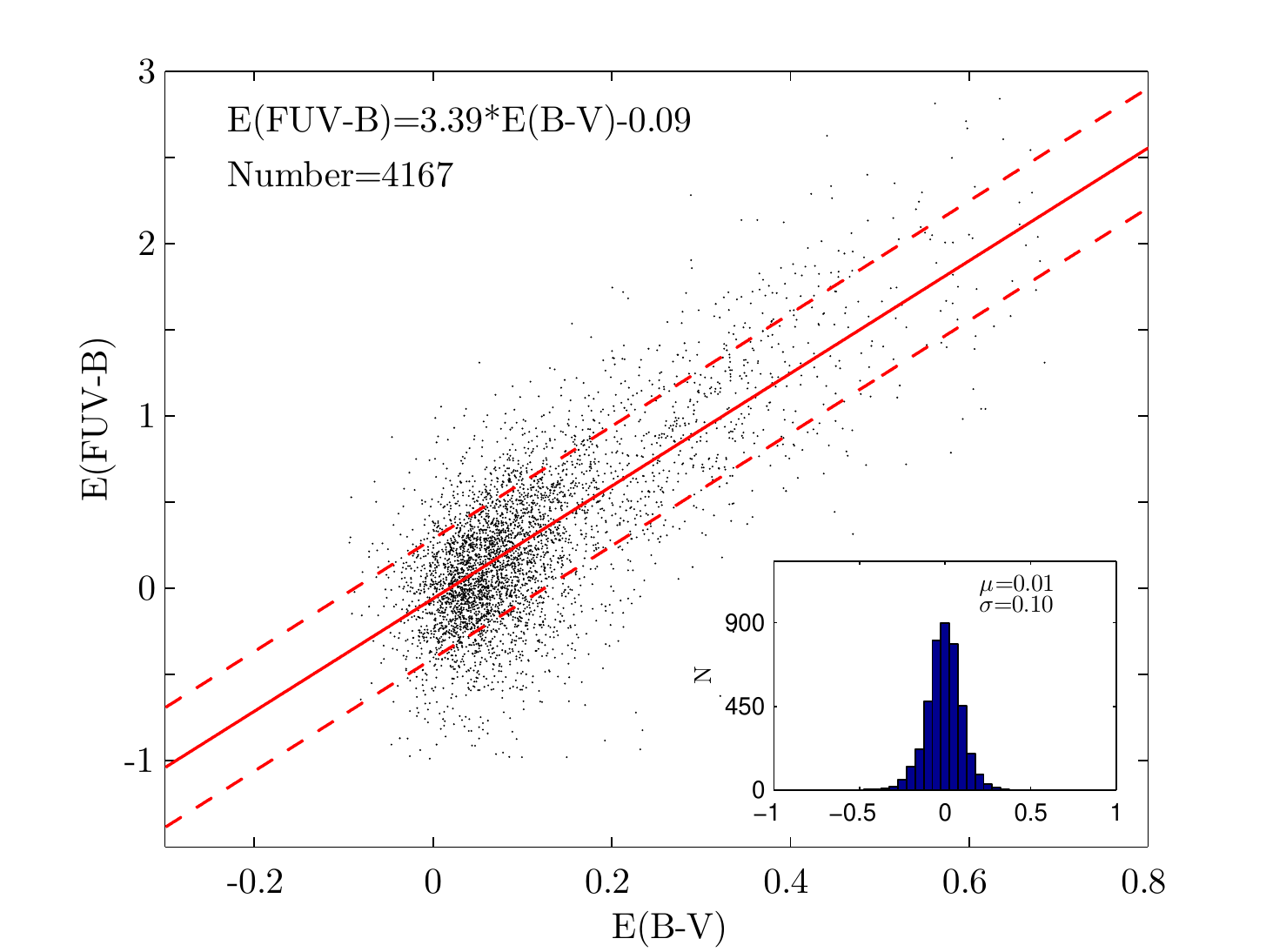}}
\caption{The same as Figure \ref{fig11}, but for the color excesses, $E_{{\rm FUV,B}}$ and $E_{{\rm B,V}}$.
\label{fig12}}
\end{figure}

The robust linear fitting  yields $E_{{\rm NUV,B}}$/$E_{{\rm B,V}}=3.77$ and $E_{{\rm FUV,B}}$/$E_{{\rm B,V}}=3.39$. This deduces $E_{{\rm FUV,NUV}}$/$E_{{\rm B,V}}=-0.38$, a more favourable value than -0.44 derived from a direct linear fitting of $E_{{\rm FUV,NUV}}$ and $E_{{\rm B,V}}$ because $E_{{\rm FUV,NUV}}$ is very small as indicated in Figure \ref{fig5}. The residual of the fitting is calculated by the perpendicular distance to the fitting line, which has a dispersion of 0.05 for $E_{{\rm NUV,B}}$/$E_{{\rm B,V}}$ and 0.10 for $E_{{\rm FUV,B}}$/$E_{{\rm B,V}}$. The uncertainty of the slope is measured by the dispersion of the slopes which are obtained from linear fitting to the sub-samples grouped according to the temperature, since the slope is irrelevant to the temperature. For $E_{{\rm NUV,B}}$/$E_{{\rm B,V}}$ for which the stars have $\Teff$ from 6500\,K to 8500\,K, the sub-sample is formed with a bin of 250\,K (the mean error of $\Teff$ is about 85\,K, 250\,K is about 3 time of the mean error) and a step of 10\,K with a moving window. Thus 151 sub-samples are built. For $E_{{\rm FUV,B}}$/$E_{{\rm B,V}}$ for which the stars have $\Teff$ from 7000\,K to 8500\,K,  101 sub-samples are constructed.  The dispersion turns out to be 0.08 in $E_{{\rm NUV,B}}$/$E_{{\rm B,V}}$, and 0.17 in $E_{{\rm FUV,B}}$/$E_{{\rm B,V}}$. The uncertainty of $E_{{\rm FUV,NUV}}$/$E_{{\rm B,V}}$ is thus 0.19. The results of fitting are present in Figure~\ref{fig11}, Figure~\ref{fig12} and  Table\,\ref{table2}. It is worthy to mention that the intercept in both $NUV$ and $FUV$ case is negative. As mentioned above, the intercept is expected to be null if both color excesses are ideally determined. The non-zero intercept implies a possibly systematical shift in intrinsic color indexes. Though we have no intension to correct for this systematical shift, it should be noted that the slope could be smaller if the line is forced to pass through the origin. That means the derived color excess ratio may be smaller. Fortunately, the intercepts are rather small in comparison with the color excess $E_{{\rm NUV,B}}$ or $E_{{\rm FUV,B}}$. This analysis of uncertainty and intercepts is applicable to the following cases in which the intrinsic color indexes are based on the PARSEC model and the SFD low-extinction sightlines.

\subsubsection{From the PARSEC intrinsic colors}

\begin{figure}
\centering
\centerline{\includegraphics[scale=1]{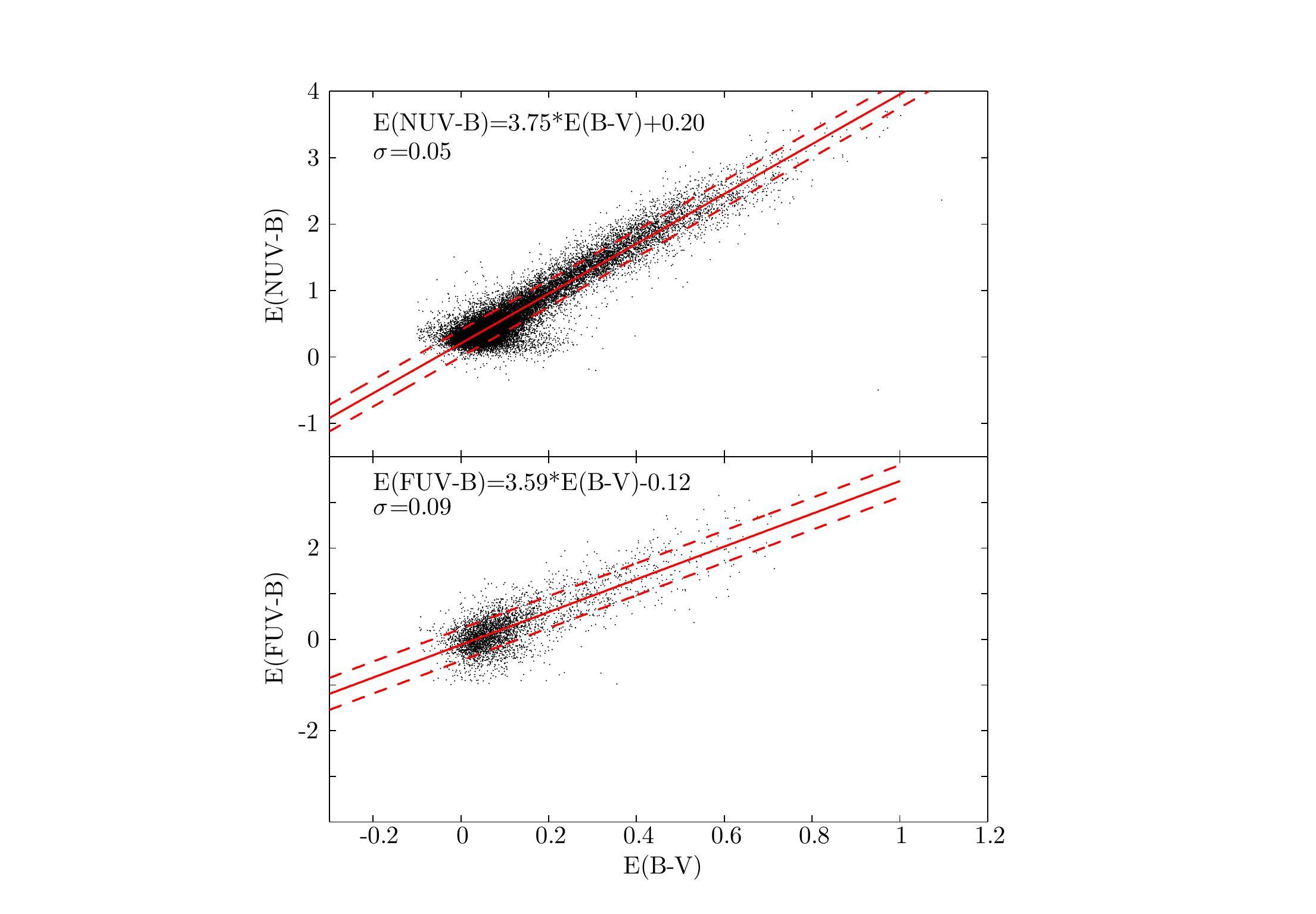}}
\caption{Linear fitting of the color excesses between $E_{{\rm NUV,B}}$ and $E_{{\rm B,V}}$, $E_{{\rm FUV,B}}$ and $E_{{\rm B,V}}$ based on the intrinsic color indexes derived from the PARSEC model (see text for details).
\label{fig13}}
\end{figure}

The intrinsic color index derived by the PARSEC stellar model differs slightly from that calculated by the blue-edge method, as discussed Section 3.4. Using the set of intrinsic color indexes from PARSEC, we also obtained a group of color excess ratios shown in Figure \ref{fig13} and Table~\ref{table2}. They agree very well in $E_{{\rm NUV,B}}$/$E_{{\rm B,V}}$, being 3.77 from the blue edge method and 3.75 from the PARSEC model. On the other hand, $E_{{\rm FUV,B}}$/$E_{{\rm B,V}}$ exhibits slight difference. The blue edge method results in $E_{{\rm FUV,B}}$/$E_{{\rm B,V}} = 3.39\pm0.17$,  while the PARSEC method results in $E_{{\rm FUV,B}}$/$E_{{\rm B,V}} = 3.59\pm0.20$. However, with the uncertainty  taken into account, this difference is fully acceptable, and the result from the PARSEC intrinsic colors is consistent with the blue-edge method.


\subsubsection{From the SFD intrinsic colors}

\begin{figure}
\centering
\centerline{\includegraphics[scale=1]{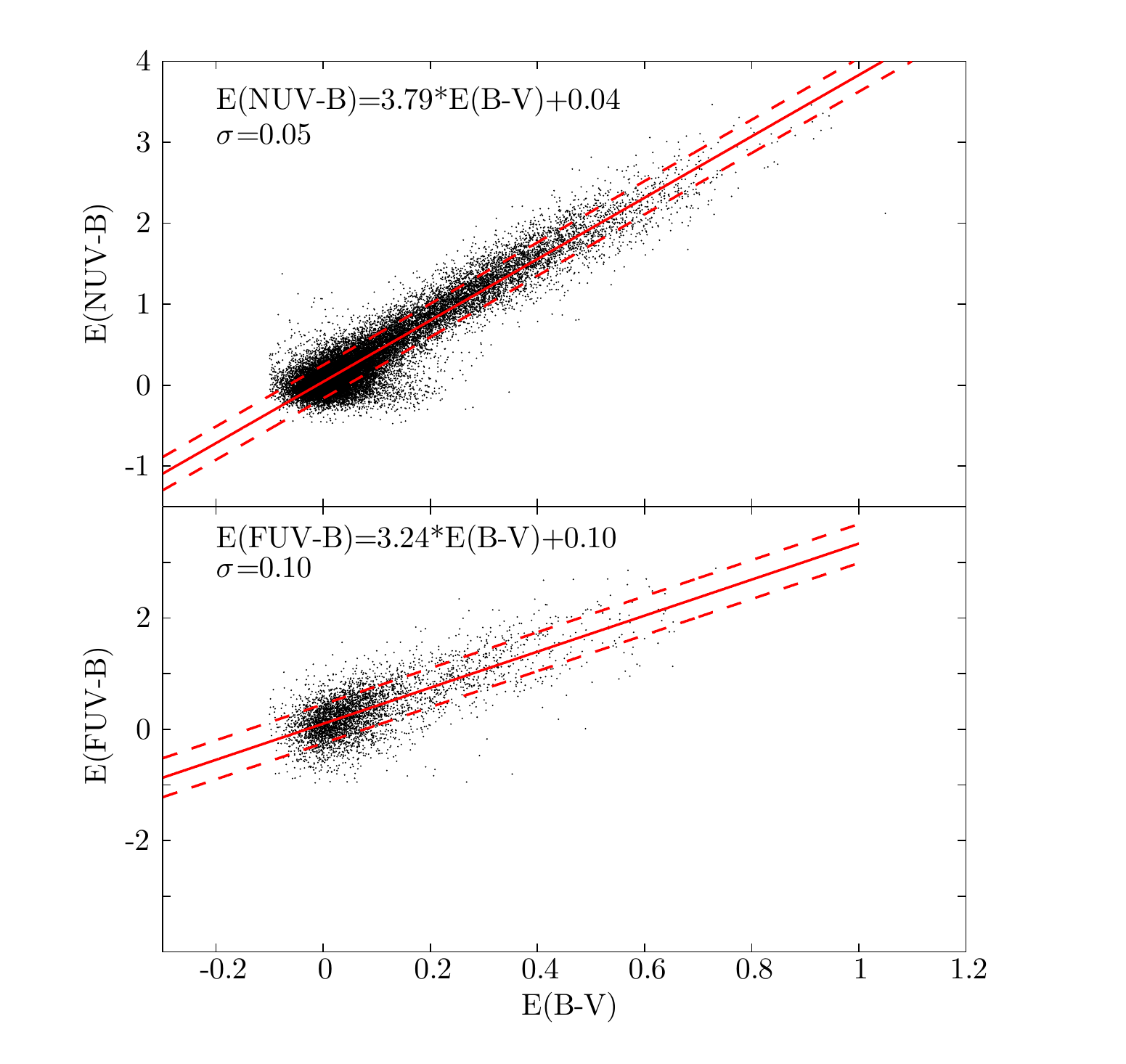}}
\caption{Linear fitting of the color excesses between $E_{{\rm NUV,B}}$ and $E_{{\rm B,V}}$, $E_{{\rm FUV,B}}$ and $E_{{\rm B,V}}$ based on the intrinsic color indexes derived from the SFD sightlines with E(B-V) $<$ 0.05 (see text for details).
\label{fig14}}
\end{figure}

The color excess ratio is also calculated for the case that the intrinsic color indexes are determined based on the sightlines with $E_{{\rm B,V\_SFD}} <0.05$ as described in Section 3.5. By using the same method as the blue-edge case, the derived ratios and uncertainties are
$E_{{\rm NUV,B}}/E_{{\rm B,V}}=3.79\pm0.08$, $E_{{\rm FUV,B}}$/$E_{{\rm B,V}}=3.24\pm0.16$ and $E_{{\rm FUV,NUV}}$/$E_{{\rm B,V}}=-0.55\pm0.18$. The results are present in Figure \ref{fig14}. With the uncertainty taken into account, these values are very consistent with that from the blue-edge method. In addition, that $E_{{\rm NUV,B}}/E_{{\rm B,V}} > E_{{\rm FUV,B}}/E_{{\rm B,V}}$ agrees with the blue-edge method, which can be explained by the enhancement to the $NUV$ band by the 2175${\rm \AA}$ bump.   This result is expected since the slope of the linear fitting is not changed if only a systematic shift occurs to the axis. In a word, the result based on the SFD selected zero-reddening stars is highly consistent with the blue-edge method.

\subsubsection{Variation}

The variations in UV extinction were illustrated clearly by the IUE observations \citep{1999PASP..111...63F, 1990ApJS...72..163F}. \citet{1999PASP..111...63F} showed 80 UV extinction curves whose color excess ratio $E_{{\rm \lambda,V}}$/$E_{{\rm B,V}}$  vary by a factor up to 6-8.

\begin{figure}
\centering
\centerline{\includegraphics[scale=1]{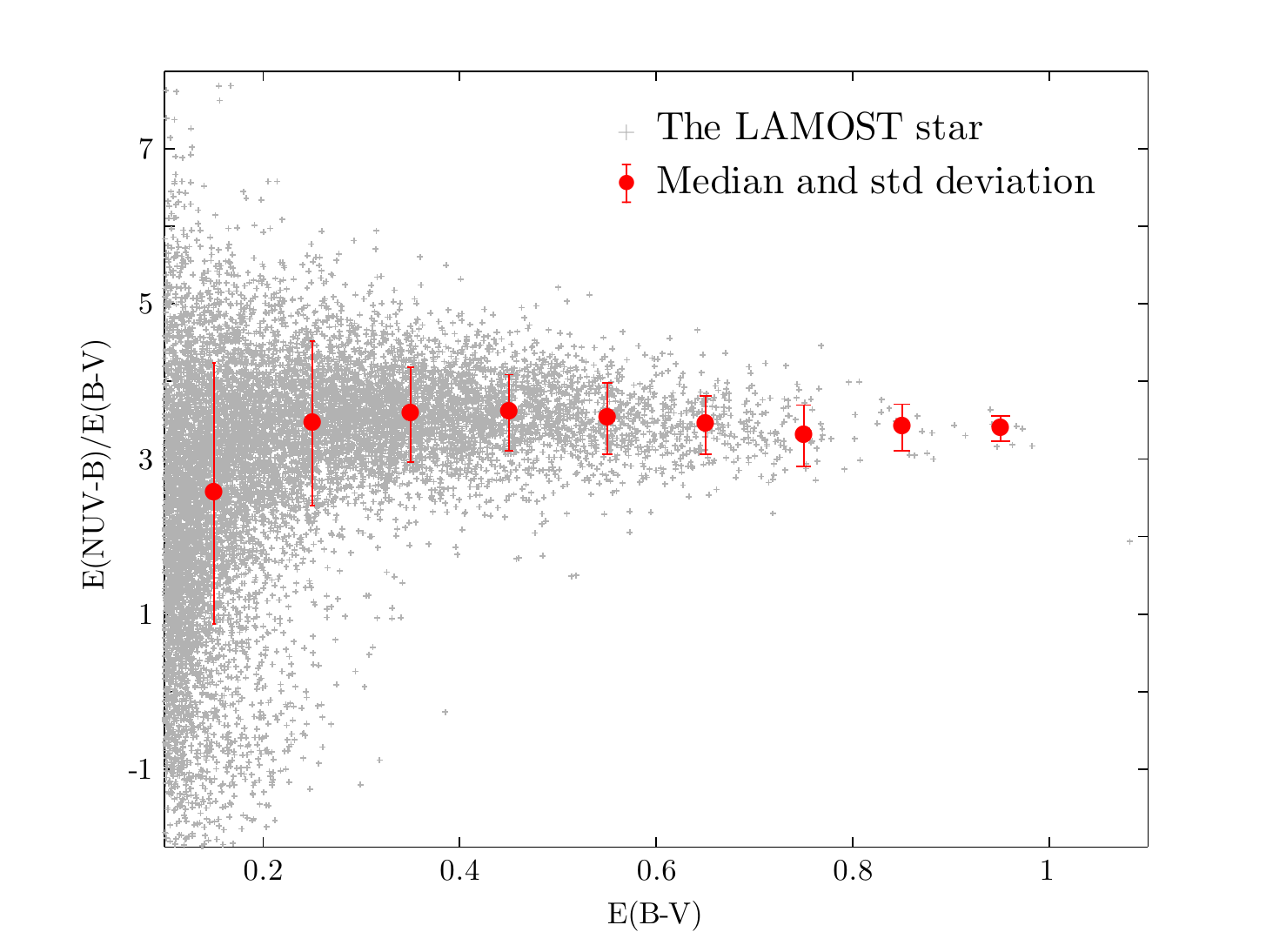}}
\caption{The distribution of E(NUV-B)/E(B-V) with E(B-V), where the grey cross denotes the ratio of individual star and the red dot denotes the mean value in a bin of 0.1 of E(B-V) with the standard deviation.
\label{fig15}}
\end{figure}

The present sample consists of numerous stars. Meanwhile its range of extinction extends only to about 1.0 mag in $E_{{\rm B,V}}$, implying pure diffuse sightlines and no dense environment. Because the extinction law is determined by the properties of interstellar dust, its variation may depend on the environment. The total-to-selective ratio $\RV$, the characteristic parameter of interstellar extinction law,  is believed to be larger in dense clouds than in diffuse medium. Whether $E_{{\rm NUV,B}}/E_{{\rm B,V}}$ varies with $E_{{\rm B,V}}$ is examined by the median of the ratio $E_{{\rm NUV,B}}/E_{{\rm B,V}}$ for individual stars within a bin of 0.1 in $E_{{\rm B,V}}$.  The stars with $E_{{\rm B,V}}<0.1$ are excluded because of their large uncertainty.   Figure \ref{fig15} displays the result, with both the median and the standard deviation, where the star with the largest $E_{{\rm B,V}}$ being about 1.1 is not taken into account for lacking of statistical significance. The median value for $E_{{\rm B,V}}$ between 0.1 and 0.2 cannot be significant since this range is smaller than three times of the error of $E_{{\rm B,V}}$ (0.064).
The color excess ratio $E_{{\rm NUV,B}}/E_{{\rm B,V}}$ decreases with $E_{{\rm B,V}}$. The amplitude of variation is small, between 3.40 and 2.85. These values are not the same as  the ratio derived from linear fitting. Because the intercept of the linear fitting is -0.13, a negative number, the ratio from linear fitting is bigger than the individual color excess ratio, which is why we did not use linear fitting to discuss the variation. If the standard deviation is counted, we can hardly claim the ratio changes. For the $FUV$ band with much fewer sources and smaller range of $E_{{\rm B,V}}$, the ratio $E_{{\rm FUV,B}}/E_{{\rm B,V}}$ has no apparent variation with $E_{{\rm B,V}}$, whose implication cannot be exaggerated.


\subsection{Comparison with previous results }

\begin{figure}
\centering
\centerline{\includegraphics[scale=1]{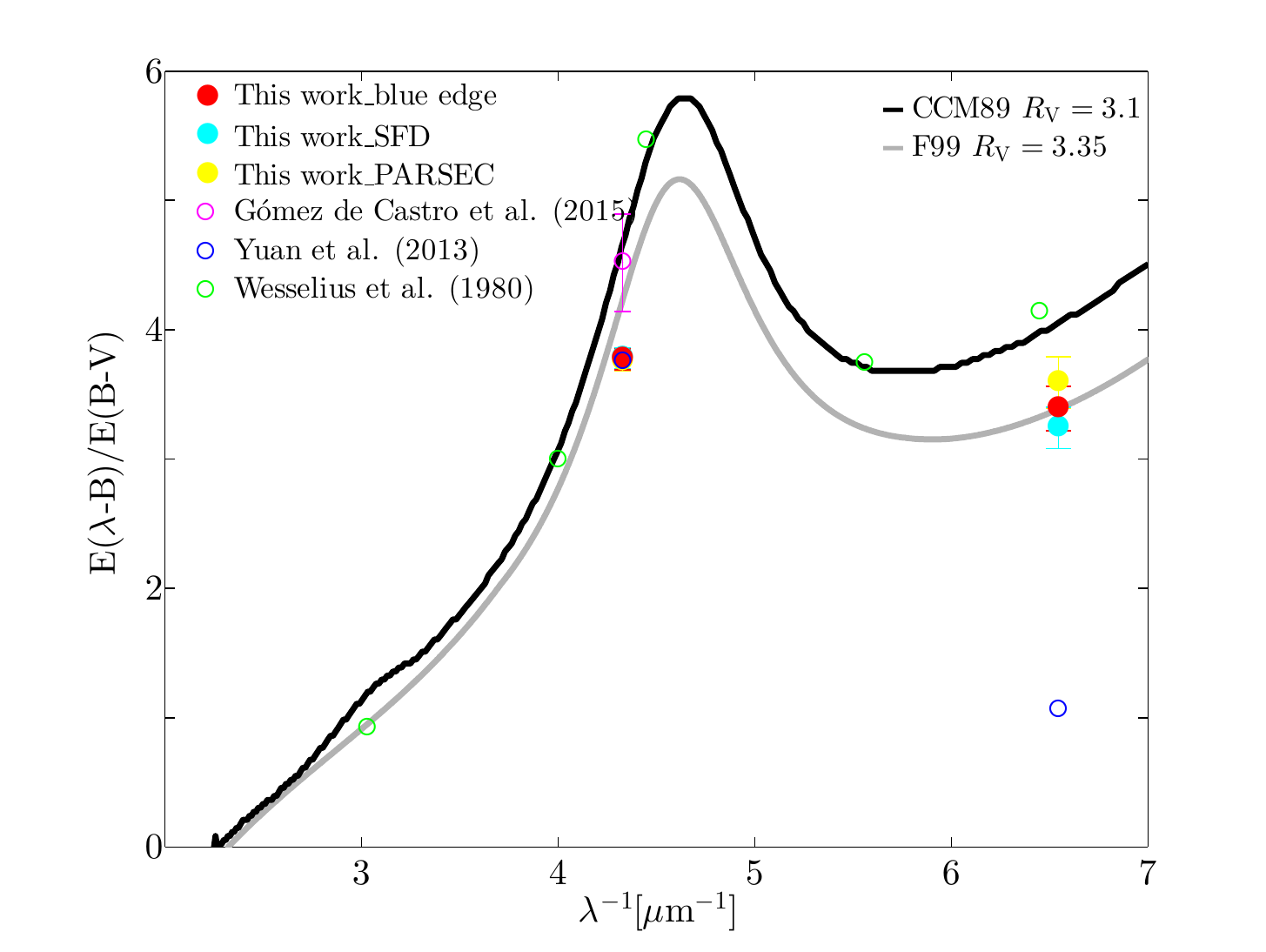}}
\caption{The color excess ratio E(NUV-B)/E(B-V) and E(FUV-B)/E(B-V) in the extinction curve with the results from other works. The black and grey lines are the extinction curve derived from the analytical formula of \citet{1989ApJ...345..245C} for $\RV=3.1$ and  \citet{1999PASP..111...63F} for $\RV=3.35$ respectively.
\label{fig16}}
\end{figure}

Our results are compared with previous works and shown in Figure~\ref{fig16} and Table\,\ref{table2}.

$E_{{\rm NUV,B}}$/$E_{{\rm B,V}}$ ($3.77 \pm 0.08$) is highly consistent with the value of $E_{{\rm NUV,g}}$/$E_{{\rm B,V}}$ (3.75) derived by \citet{2013MNRAS.430.2188Y} based on the SDSS spectroscopic survey. In comparison, \citet{2015MNRAS.449.3867G} found an apparently larger $E_{{\rm NUV,B}}$/$E_{{\rm B,V}}$ that the average value toward seven sightlines turns out to be $4.52 \pm 0.38$. However, their sightlines are all towards the Taurus-Auriga molecular complex that may be dense cloud with possibly a dust size growth. On the other hand, our sample is not biased to any specific interstellar environment while it is in fact inclined to low-extinction sightlines with $E_{{\rm B,V}} < 1.0$.  The discrepancy with \citet{2015MNRAS.449.3867G} could lie in the difference of interstellar environment. But a close check indicates that \citet{2015MNRAS.449.3867G} did not probe deep in the cloud and our study shows no apparent variation of $E_{{\rm NUV,B}}$/$E_{{\rm B,V}}$ within $E_{{\rm B,V}} < 1.0$,  the difference of methods in deriving the extinction may be the major reason. 
In comparison with the widely used extinction formulae, our $E_{{\rm NUV,B}}$/$E_{{\rm B,V}}=3.77 \pm 0.08$ is smaller than that of \citet{1989ApJ...345..245C} (4.64 at $\RV=3.1$) and close to  \citet{1999PASP..111...63F} (4.23 at $\RV=3.35$). This is not surprising since these analytical formulae were based mainly on the results from the IUE spectroscopy and the ANS survey (e.g.\citet{1980A&A....85..221W} shown by green circle in Figure~\ref{fig16}).

$E_{{\rm FUV,B}}$/$E_{{\rm B,V}}$ ($3.39 \pm 0.17$) is significantly larger than 1.06 derived by \citet{2013MNRAS.430.2188Y}. In fact, \citet{2013MNRAS.430.2188Y} did not calculate the value of $E_{{\rm FUV,B}}$/$E_{{\rm B,V}}$, and this value of 1.06 is inferred from their values of $E_{{\rm NUV,g}}$/$E_{{\rm B,V}}$ and $E_{{\rm FUV,NUV}}$/$E_{{\rm B,V}}$ and may suffer large uncertainty. On the other hand, this value is smaller than the value by \citet{1980A&A....85..221W}  and the value of 4.15 from \citet{1989ApJ...345..245C}. The number 3.39 closely matches the point on the curve of \citet{1999PASP..111...63F} with $\RV=3.35$.

$E_{{\rm FUV,NUV}}$/$E_{{\rm B,V}}$ is equal to -0.38. This negative value indicates that the $NUV$ band has higher extinction than the $FUV$ band due to the contribution of the 2175${\rm \AA}$ bump. This agrees with the tendency of the \citet{1989ApJ...345..245C} and \citet{1999PASP..111...63F} analytical formula.

%

\section{Summary}

The average UV extinction law is derived
with high precision in the two \emph{GALEX} bands
with reference to the optical $B$ and $V$ bands,
for the entire Galaxy by
using A- and F-type dwarf stars
as the extinction tracer.
This derivation is based on
the color-excess method,
with the intrinsic stellar color indices
determined from the stellar spectra
obtained by the LAMOST survey.

The major results of this paper are as follows:
\begin{enumerate}
\item An analytical relation of stellar intrinsic colors
      $C^{0}_{\rm B,V}$,$C^{0}_{\rm NUV,B}$, $C^{0}_{\rm FUV,B}$, $C^{0}_{\rm FUV,NUV}$
      is determined for the dwarf stars in the LAMOST survey with
      the stellar effective temperature $\Teff$
      between $\sim$6500--8500\,K
      for given metallicity ranges (-0.625 -- 0.125). The UV intrinsic colors depend strongly on the metallicity range, changing by about 0.3 mag in $C^{0}_{\rm NUV,B}$, $C^{0}_{\rm FUV,B}$ with $Z$ from -0.5 to 0.0.
%
\item The color excess $E_{{\rm NUV,B}}$ is calculated for 25,496, and $E_{{\rm FUV,B}}$ and $E_{{\rm FUV,NUV}}$ for 4,255 A- and F-type dwarfs stars from the LAMOST survey.
\item The mean color excess ratios are derived to be
      $E_{{\rm NUV,B}}$/$E_{{\rm B,V}} \approx3.77\pm0.08$,
      $E_{{\rm FUV,B}}$/$E_{{\rm B,V}} \approx3.39\pm0.17$,
      $E_{{\rm FUV,NUV}}$/$E_{{\rm B,V}} \approx-0.38\pm0.19$.
      $E_{{\rm NUV,B}}$/$E_{{\rm B,V}}$ agree very well with previous results. 
      In general, they are smaller than the corresponding values of the extinction curve at $\RV=3.1$ of \citet{1989ApJ...345..245C}, and  agree well with the extinction curve derived by \citet{1999PASP..111...63F} for $\RV=3.35$.
%

\end{enumerate}


\acknowledgments{We thank Li Chen, Jun~Li, Jiaming~Liu
                  and Mengyao~Xue for their discussions. Special thanks go to the anonymous referee for his/her very detailed and helpful suggestions, which improved the paper significantly. This work is supported by NSFC through Projects 11533002 and U1631104, and the 973 Program 2014CB845702. This work made use of the data taken by \emph{GALEX}, LAMOST and APASS.
}


\facilities{\emph{GALEX}, LAMOST, APASS}

\clearpage

\begin{table}
\begin{center}
\caption{\label{table1}The coefficients of Equation (1)}
\vspace{0.05in}
\begin{tabular}{|l|c|c|c|c|c|c|c|c|c|c|c|c|}
\tableline
$C^{0}_{\rm \lambda1,\lambda2}$  & Z & $a_{0}$&$a_{1}$&$a_{2}$ \\
\tableline
$C^{0}_{\rm B,V}$           & [-0.625,-0.375]&2.56&-0.49&0.024\\
\tableline
$C^{0}_{\rm B,V}$           & [-0.375,-0.125]&2.79&-0.52&0.024 \\
\tableline
$C^{0}_{\rm B,V}$           & [-0.125, 0.125] &3.35&-0.66&0.033\\
\tableline
$C^{0}_{\rm NUV,B}$         & [-0.625,-0.375]&14.17&-2.76&0.16 \\
\tableline
$C^{0}_{\rm NUV,B}$         & [-0.375,-0.125]&10.67&-1.68&0.083\\
\tableline
$C^{0}_{\rm NUV,B}$         & [-0.125, 0.125]&14.17&-2.50&0.13\\
\tableline
$C^{0}_{\rm FUV,B}$         &[-0.625,-0.375]&13.88&-0.53&-0.071\\
\tableline
$C^{0}_{\rm FUV,B}$         & [-0.375,-0.125] &17.11&-1.09& -0.049\\
\tableline
$C^{0}_{\rm FUV,B}$         & [-0.125, 0.125]&32.71&-5.13&0.22\\
\tableline
$C^{0}_{\rm FUV,NUV}$       & [-0.625,-0.375]&-1.45&2.55&-0.25\\
\tableline
$C^{0}_{\rm FUV,NUV}$       &[-0.375,-0.125] &7.31&0.38&-0.12\\
\tableline
$C^{0}_{\rm FUV,NUV}$       & [-0.125, 0.125] &17.00&-2.20&0.054\\
\tableline
\end{tabular}
\end{center}
\end{table}


\begin{table}
\begin{center}
\caption{\label{error_c0} Uncertainty and source of the intrinsic color indexes}
\vspace{0.05in}
\begin{tabular}{|l|c|c|c|c|c|c|c|}
\tableline \tableline
            & $C^{0}_{\rm B,V}$ & $C^{0}_{\rm NUV,B}$ & $C^{0}_{\rm FUV,B}$ & $C^{0}_{\rm FUV,NUV}$ \\
\tableline
  From $\Teff$    & 0.016 &0.049 & 0.123 & 0.094\\
\tableline
  From metallicity     & 0.009 & 0.104 &0.132&0.050 \\
\tableline
  From the blue-edge method    & 0.013 & 0.011 &0.022 &0.010 \\
\tableline
From photometry & 0.042 & 0.048 & 0.150 & 0.150 \\

\tableline
  Total uncertainty                & 0.048 & 0.136 &0.242 &0.187 \\
\tableline
\end{tabular}
\end{center}
\end{table}

\begin{table}
\begin{center}
\caption{\label{table2}The color excess ratios}
\vspace{0.05in}
\setlength{\tabcolsep}{0.5mm}
\begin{tabular}{|l|c|c|c|c|c|c|c|c|}
\tableline
                  &\multicolumn{3}{c|}{This\ work}& Y2013{\tablenotemark{a}} & G2015{\tablenotemark{b}} &CCM89{\tablenotemark{c}} & F99{\tablenotemark{d}} \\
\tableline
Method for $C^{0}_{\rm \lambda1,\lambda2}$   &BlueEdge&PARSEC& E(B-V)$<=$0.05(SFD)&$-$&$-$&$-$&$-$\\
\tableline
  $E_{{\rm NUV,B}}$/$E_{{\rm B,V}}$& \textbf{$3.77\pm0.08$} &$3.75\pm0.07$& $3.79\pm0.08$&3.75{\tablenotemark{e}}&4.52&4.64&4.23\\
\tableline
 $E_{{\rm FUV,B}}$/$E_{{\rm B,V}}$& \textbf{$3.39\pm0.17$} & $3.59\pm0.20$& $3.24\pm0.16$&1.06{\tablenotemark{f}}&$-$&4.05&3.39\\
\tableline
  $E_{{\rm FUV,NUV}}$/$E_{{\rm B,V}}$ &\textbf{$-0.38\pm0.19$}{\tablenotemark{g}} &$-0.16\pm0.21${\tablenotemark{g}}& $-0.55\pm0.18${\tablenotemark{g}} &-2.69&$-$&-0.59{\tablenotemark{g}}&-0.84{\tablenotemark{g}}\\
\tableline
\end{tabular}
\end{center}
\tablenotetext{a}{\citet{2013MNRAS.430.2188Y}}
\tablenotetext{b}{\citet{2015MNRAS.449.3867G}}
\tablenotetext{c}{\citet{1989ApJ...345..245C} with $\RV$=3.1}
\tablenotetext{d}{\citet{1999PASP..111...63F} with $\RV$=3.35}
\tablenotetext{e}{The value of $E_{{\rm NUV,g}}$/$E_{{\rm B,V}}$ instead of $E_{{\rm NUV,B}}$/$E_{{\rm B,V}}$ }
\tablenotetext{f}{The value of $E_{{\rm FUV,g}}$/$E_{{\rm B,V}}$ instead of $E_{{\rm FUV,B}}$/$E_{{\rm B,V}}$, and calculated from $E_{{\rm NUV,g}}$/$E_{{\rm B,V}} + E_{{\rm FUV,NUV}}$/$E_{{\rm B,V}}$ }
\tablenotetext{g}{Calculated from $E_{{\rm FUV,B}}$/$E_{{\rm B,V}} - E_{{\rm NUV,B}}$/$E_{{\rm B,V}}$ }
\end{table}

\clearpage

\clearpage

\clearpage

\end{document}